\documentclass[12pt,letterpaper,twoside]{article}

\usepackage{amsmath,bm}
\usepackage{amssymb}
\usepackage{graphicx}
\usepackage{cite}
\usepackage{hyperref}
\usepackage[usenames, dvipsnames]{color}
\usepackage[font=small,labelfont=bf]{caption}
\usepackage{siunitx}
\usepackage{enumitem}
\usepackage{calc}
\usepackage[margin=3cm]{geometry}
\usepackage[section]{placeins}

\newcommand{\td}[2]{\frac{d #1}{d #2}}
\newcommand{\tn}[1]{\textnormal{#1}}
\newcommand{\mb}[1]{\mathbf{#1}}

\usepackage{tikz}
\usetikzlibrary{calc}

\begin{document}

\title{Magnetic flux pumping in 3D nonlinear magnetohydrodynamic simulations}
\author{I.~Krebs$^1$\footnote{ikrebs@pppl.gov}, S.C.~Jardin$^2$, S.~G\"unter$^3$, K.~Lackner$^3$,\\ M.~Hoelzl$^3$, E.~Strumberger$^3$, N.~Ferraro$^2$ \\\\
$^1$ Max-Planck/Princeton Research Center for Plasma Physics\\
$^2$ Princeton Plasma Physics Laboratory, Princeton, NJ, USA\\
$^3$ Max Planck Institute for Plasma Physics, Garching, Germany}

\maketitle
\abstract{A self-regulating magnetic flux pumping mechanism in tokamaks that maintains the core safety factor at $q\approx 1$, thus preventing sawteeth, is analyzed in nonlinear 3D magnetohydrodynamic simulations using the M3D-C$^1$ code. In these simulations, the most important mechanism responsible for the flux pumping is that a saturated $(m=1,n=1)$ quasi-interchange instability generates an effective negative loop voltage in the plasma center via a dynamo effect. It is shown that sawtoothing is prevented in the simulations if $\beta$ is sufficiently high to provide the necessary drive for the $(m=1,n=1)$ instability that generates the dynamo loop voltage. The necessary amount of dynamo loop voltage is determined by the tendency of the current density profile to centrally peak which, in our simulations, is controlled by the peakedness of the applied heat source profile.} 

\bigskip
\setlength{\parskip}{5pt}
\setlength{\parsep}{5pt}
\setlength{\headsep}{0pt}
\setlength{\topskip}{0pt}
\setlength{\topmargin}{0pt}
\setlength{\topsep}{0pt}
\setlength{\partopsep}{0pt}

The sawtooth instability is a relaxation-oscillation of the core plasma ubiquitous in most large tokamaks \cite{Hastie1997}. A sawtooth cycle consists in a slow rise and a subsequent fast flattening of the pressure in a central region of the plasma. A basic model that is commonly used as a point of reference for the theoretical treatment of the sawtooth instability has been proposed by Kadomtsev \cite{Kadomtsev1975}. According to this model, the sawtooth instability is caused by an internal kink instability with a $(m=1,n=1)$ mode structure, where $m$ and $n$ are the poloidal and toroidal mode number, respectively. It is destabilized by a peaking of the central toroidal current density profile that corresponds to a value of the safety factor ($q$) on the magnetic axis below unity. A $(m=1,n=1)$ magnetic island develops on the $q=1$ surface, and the magnetic reconnection proceeds until the island has entirely replaced the original plasma core, leaving the plasma center in an axisymmetric state with a flat safety factor profile close to unity. While numerical simulations have confirmed that the dynamics described by Kadomtsev's model occur within the resistive magnetohydrodynamic (MHD) model, the model does not account for a variety of experimental observations \cite{Hastie1997}. One example is that measurements on several tokamaks suggest that the magnetic reconnection process does not complete and the value of the safety factor on axis ($q_0$) stays below unity during the entire sawtooth cycle, e.g. \cite{TEXTOR1989,Blum1990,Yamada1994,Letsch2002}.

Sawteeth can provide seed islands for neoclassical tearing modes (NTMs), an instability that has the potential to degrade the energy and particle confinement of a discharge and can even lead to a disruption \cite{Chapman2010}. A mode of tokamak operation that avoids sawtoothing is the Hybrid scenario. Hybrid discharges are characterized by a flat, sometimes slightly reversed, central safety factor profile close to unity and exhibit good core confinement \cite{Stober2007}. Originally called Improved H-mode, the scenario became known as Hybrid as it represents a mode of operation in between the inductive Standard H-mode which has positive magnetic shear with $q_0 \lesssim 1$ (and exhibits sawteeth) and the fully non-inductive Steady-state scenario with reversed magnetic shear and $q_0 > 1$ \cite{Sips2005}.

Hybrid discharges have been generated in most large tokamaks \cite{Sips2002,Joffrin2005,Ide2005,Petty2015,Bock2017} by additional heating during the current ramp-up phase, which leads to broader current density profiles as a result of reduced current diffusion. Transport simulations for discharges of this type often predict $q_0<1$ during the stationary phase \cite{Gruber1999,Staebler2005,Petty2009}, which would lead to sawtoothing. However, measurements show that the toroidal current density is redistributed such that the central safety factor profile is clamped to values close to unity. The mechanism responsible for this is referred to as magnetic flux pumping \cite{Petty2009}. The self-regulating redistribution of current has the advantage that external current drive can be applied in the plasma center where it is most effective. This and the favorable confinement and stability properties make the Hybrid scenario a candidate for an Advanced Tokamak scenario \cite{Petty2015}. In order to extrapolate the properties and accessibility of the Hybrid scenario to larger future devices like ITER, it is crucial to understand in detail the mechanism behind flux pumping.

This paper aims at broadening the understanding of the flux pumping mechanism based on the explanation given in \cite{Jardin2015}. We present the asymptotic states of 3D nonlinear resistive MHD simulations that can be classified into two basic types: Sawtoothing cases where $q_0$ decreases to values significantly below unity and $q_0\approx 1$ is periodically restored by a change of the magnetic topology, and steady-state sawtooth-free cases with a helically perturbed core, low magnetic shear in the center and a central safety factor profile that stays close to unity. Tokamak plasma states similar to this stationary state have been previously examined by means of nonlinear MHD simulations \cite{Denton1987,Lukin2008,Halpern2010,Breslau2011} as well as 3D MHD equilibrium calculations \cite{Cooper2010,Cooper2011,Strumberger2014}. The results of 3D ideal MHD equilibrium calculations have been compared to linear and nonlinear ideal MHD stability calculations in \cite{Brunetti2012,Brunetti2014a}. The purpose of this work is to give an explanation of how and under which conditions such a state sustains itself nonlinearly and how this explanation can be applied to flux pumping in Hybrid discharges. 

While in the following we will discuss all types of long-term behavior that are obtained in the simulations, we focus on reviewing and elaborating on the explanation of the current redistribution mechanism (Section~\ref{mechanism}) and analyzing under which conditions this mechanism is strong enough to prevent sawtoothing in the simulations (Section~\ref{conditions}). The set-up of the presented 3D nonlinear simulations is briefly described Section~\ref{setup}. In Section~\ref{linear}, some aspects of the discussed flux pumping mechanism are analyzed in more detail by means of a linear stability analysis of an equilibrium with low central magnetic shear and $q_0\approx 1$.

\section{Simulation set-up}
\label{setup}
The presented calculations have been performed using the high-order finite element MHD code M3D-C$^1$ \cite{Jardin2012}. It uses a tensor product of reduced quintic finite elements \cite{Strang1973,Jardin2004} in the poloidal plane and Hermite cubic finite elements \cite{Strang1973} in the toroidal direction. A split-implicit time advance allows long-time integrations. The code offers several modes of operation, physics models and geometries. The full set of equations that are solved is described in \cite{Jardin2012}. In this study, we use the resistive single-fluid MHD model in toroidal geometry (see Appendix). The number of toroidal elements is eight, and each $R$-$Z$ plane has about 1000 nodes. For each of the 3D nonlinear simulations a corresponding 2D axisymmetric nonlinear calculation is done for comparison. Select 2D and 3D runs have been made at double the spatial resolution with no qualitative difference in results.

As we focus on the asymptotic states of the simulations, they cover time spans of a few $10^5 \tau_{A}$ where $\tau_{A}= l_0 \sqrt{\mu_0 \rho_0}/B_0 \approx 0.3\, \mu \tn{s}$ is the Alfv\'en time and $l_0$, $\rho_0$ and $B_0$ are typical values for the length scale, mass density and magnetic field, respectively. The simulations effectively start from a safety factor profile that is flat and close to unity in the plasma core which is generated by an initial sawtooth reconnection event. We analyze a large set of calculations, obtained by varying three parameters: the poloidal $\beta$, the perpendicular heat diffusion coefficient $\chi_\perp$ together with the strength of the heat source, and the peakedness of the heat source profile. In all cases, the simulations are set up such that, in the absence of instabilities, the heat source would drive the central safety factor profile to a value below unity.

All other parameters are held fixed as described in \cite{Jardin2015}. To keep the Spitzer resistivity similar between simulations with different values of $\beta$, it is rescaled accordingly. The central value of the resistivity of $\eta \approx 4 \cdot 10^{-6} \, \Omega \tn{m}$ is a factor of $10^2 .. 10^3$ higher than realistic resistivities in modern large tokamaks due to limited computational resources. To ensure a realistic ratio of resistive and heat diffusion time scales, the perpendicular heat diffusion coefficient is scaled up similarly. In order to examine the influence of varying the resistivity while keeping its ratio to the perpendicular heat diffusion coefficient fixed, one simulation (case \lq n\rq) has been rerun with both $\eta$ as well as $\chi_\perp$ and the strength of the heat source scaled up by a factor of three (case \lq nX3\rq). More details on the simulation set-up are given in the Appendix.

\section{Flux pumping mechanism}
\label{mechanism}
As described in \cite{Jardin2015} we find an asymptotic state in 3D nonlinear MHD simulations which is characterized by a central region with very low magnetic shear where the safety factor profile has a value close to unity and which is thus stable to the internal kink instability. This state features a stationary $(m=1,n=1)$ perturbation in the core, in particular a $(m=1,n=1)$ helical flow as shown in Figure~\ref{flow}. The flow is generated by a saturated quasi-interchange instability, a pressure-driven $(m=1,n=1)$ instability allowed for by the ultra-low magnetic shear \cite{Hastie1988,Waelbroeck1988,Waelbroeck1989}. The comparison of the safety factor profiles in such a simulation and in a corresponding 2D axisymmetric calculation in Figure~\ref{3D2Dq} shows that a 3D effect is responsible for the observed flattening of the central current density. The 2D simulation can be seen as an analog to the transport simulations in \cite{Gruber1999,Staebler2005,Petty2009} which falsely predict $q_0<1$ for the described Hybrid discharges.
\begin{figure}
  \centering
  \begin{tikzpicture}[remember picture,align=center] \small
  \node(im){\includegraphics[width=0.5\textwidth]{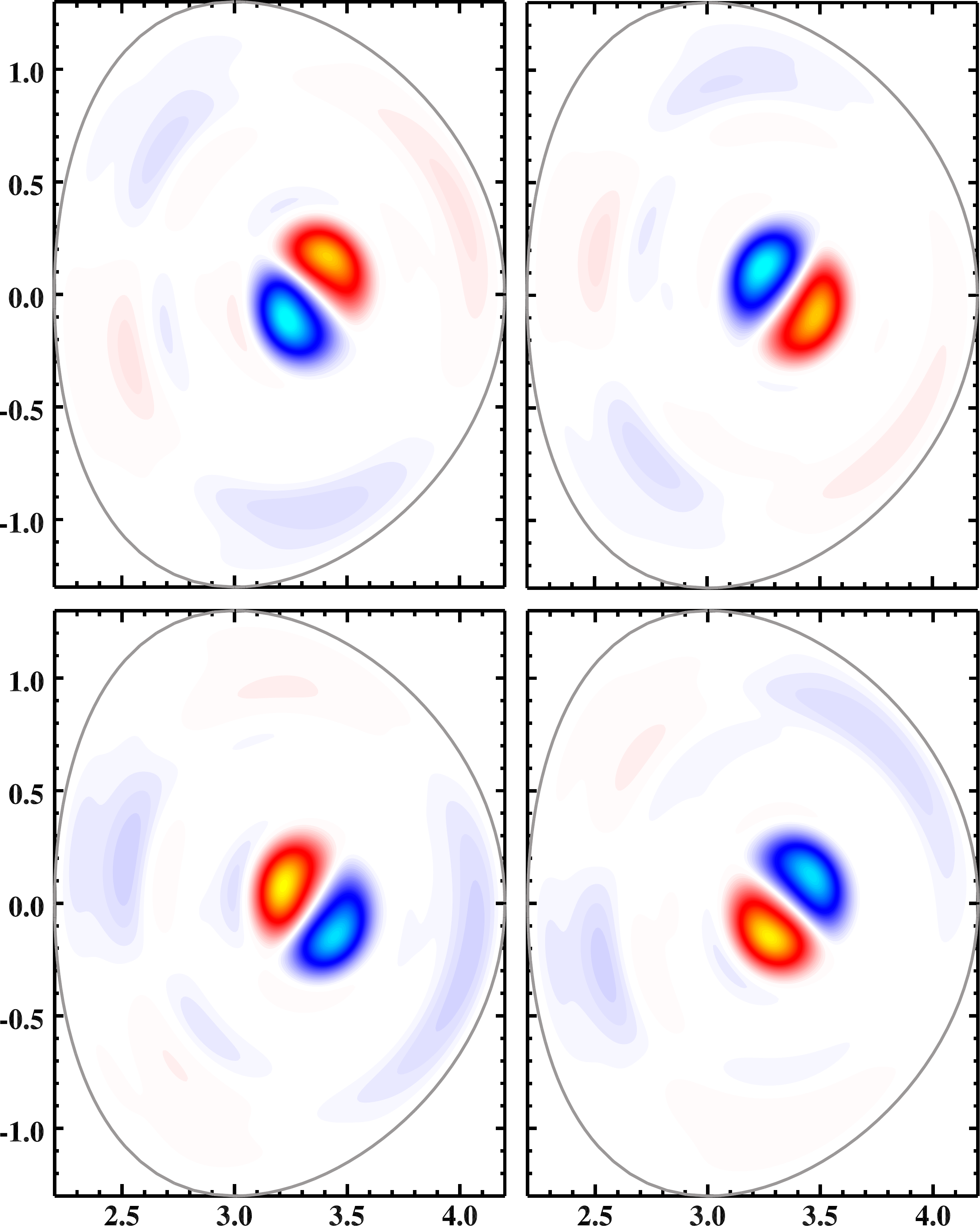}};
  \node[anchor=north] at (im.south) {\hspace{0.5cm}$R [m] \qquad\qquad\qquad\qquad$\hspace{0.2cm}$ R [m]$};
  \node[anchor=south,rotate=90] at (im.west) {\hspace{0.5cm}$ Z [m] \qquad\qquad\qquad\qquad\qquad$\hspace{0.4cm}$ Z [m]$};
  \node[anchor=south east] at ($(im.center)+(4pt,8pt)$) {$0^\circ$};
  \node[anchor=south east] at ($(im.east)+(-3pt,8pt)$) {$90^\circ$};
  \node[anchor=south east] at ($(im.south)+(4pt,12pt)$) {$270^\circ$};
  \node[anchor=south east] at ($(im.south east)+(-3pt,12pt)$) {$180^\circ$};  
  \end{tikzpicture}
  \caption{Difference between the poloidal velocity stream function in a stationary 3D simulation and in the corresponding 2D axisymmetric calculation for different toroidal angles. Negative values are indicated in blue and positive values in red. It can be seen that the velocity perturbation has the form of a $(m=1,n=1)$ convection cell in the plasma center. (Case \lq n\rq). Note that M3D-C$^1$ uses a form of a Helmholtz representation for the poloidal velocity field \cite{Ferraro2009}. The incompressible component, whose stream function is plotted here, greatly exceeds the compressible component of the field.}
  \label{flow} 
\end{figure}
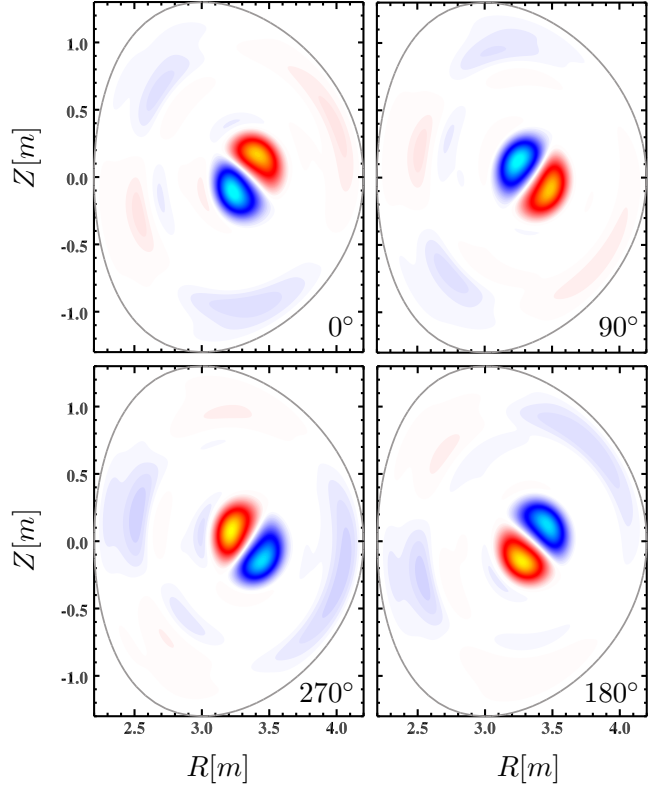
\begin{figure}
  \centering
  \includegraphics[width=0.6\textwidth]{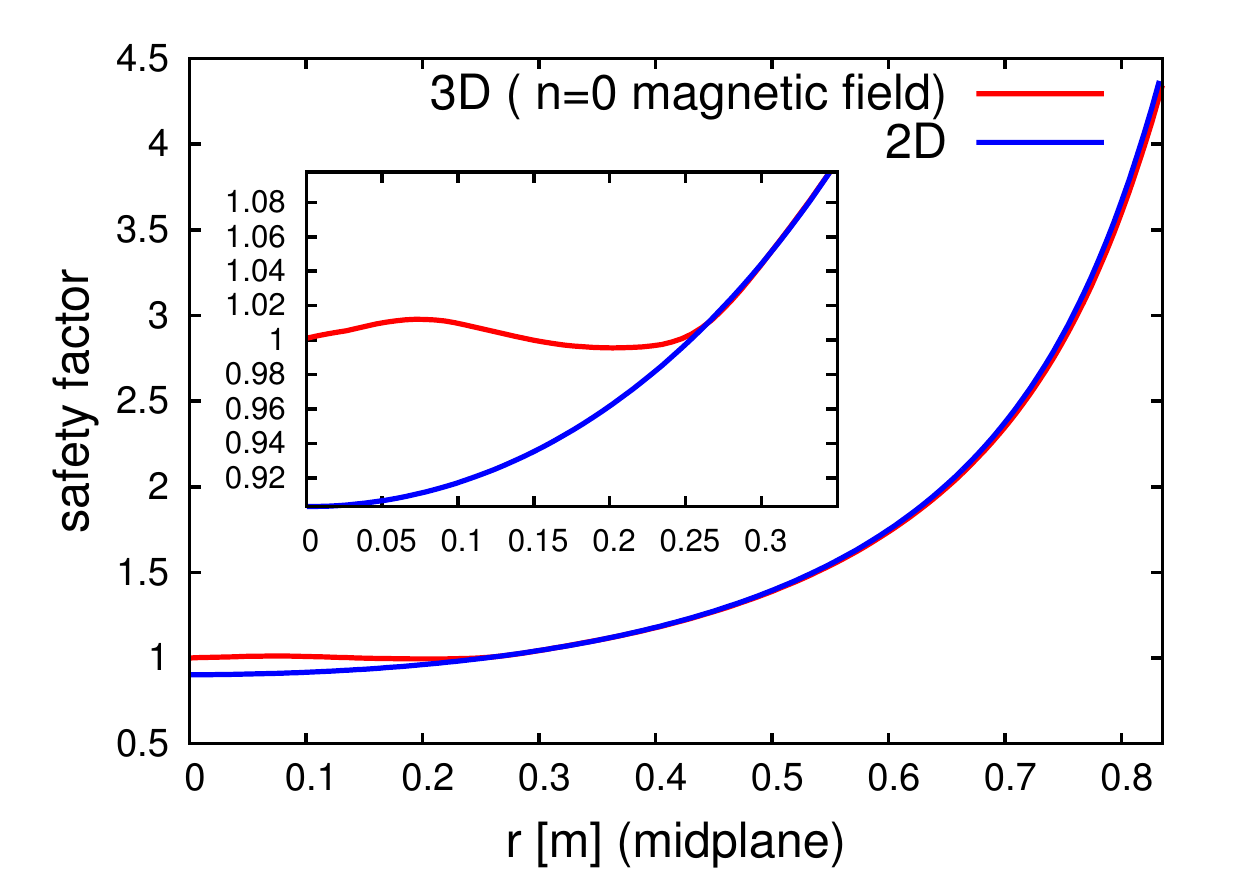}
  \caption{Comparison of the safety factor profile in the asymptotic state of a 3D and the corresponding 2D axisymmetric simulation. For the 3D case, the safety factor profile has been calculated using the toroidally averaged magnetic field. (Case \lq n\rq).}
  \label{3D2Dq} 
\end{figure}

In the following it is illustrated which 3D effects can alter the background ($n=0$) toroidal current density by analyzing the induction equation
\begin{align}
\partial_t \mathbf{B} = - \nabla \times \mathbf{E}
\label{induction1} 
\end{align}
where $\mathbf{E}$ is the electric field. After replacing the magnetic field $\mathbf{B}$ by $\mathbf{B}=\nabla \times \mathbf{A}$, where $\mathbf{A}$ is the magnetic vector potential, integration of Equation~\eqref{induction1} leads to
\begin{align}
\partial_t \mathbf{A}= -\mathbf{E}-\nabla \Phi \,.
\label{induction2}
\end{align}
Here $\Phi$ is a single valued potential. Using cylindrical coordinates ($R,\phi,Z$), the projection of Equation~\eqref{induction2} onto the toroidal direction yields
\begin{align}
\partial_t \Psi = -R\eta J_{\phi} +R\hat{\phi} \cdot \left(  \mathbf{v} \times \mathbf{B} \right) -R \hat{\phi} \cdot \nabla \Phi \,.
\label{induction3}
\end{align}
Here Ohm's law $\mathbf{E}=\eta \mathbf{J}-\mathbf{v} \times \mathbf{B}$ has been used to eliminate the electric field, and the toroidal component of the magnetic vector potential has been replaced by $\Psi \nabla \phi$, where $\Psi$ denotes the negative of the poloidal magnetic flux per radian. We split all quantities into an axisymmetric part\footnote{Note, that the index 0 always refers to the $n=0$ component of the indexed quantity, with the exception of $q_0$ which refers to the value of the safety factor on axis to be consistent with the common notation.} and a non-axisymmetric part and only take into account the dominant $n=1$ component of the latter. The toroidal average of Equation~\eqref{induction3} then gives
\begin{align}
\partial_t \Psi_0 = -R\eta_0 J_{\phi,0} - R \left[ \eta_1 J_{\phi,1} \right]_{n=0} + R \left[ \hat{\phi} \cdot \left( \mathbf{v}_1 \times \mathbf{B}_1 \right) \right]_{n=0} \, .
\label{induction4}
\end{align}
Note that $\nabla \Phi_0=0$ and we assume $\mathbf{v}_0=0$. The $n=0$ quantities can be expressed in terms of the corresponding quantities in the 2D axisymmetric system plus a deviation due to the influence of the 3D perturbation on the $n=0$ background: $\Psi_0 = \Psi_{2D} + \Delta \Psi$, $\eta_0 =\eta_{2D} + \Delta \eta$ and $J_{\phi,0}= J_{\phi,2D} + \Delta J_{\phi}$. The toroidal induction equation for the 2D system reads
\begin{align}
\partial_t \Psi_{2D} = -R \eta_{2D} J_{\phi,2D} \,.
\label{induction2D}
\end{align}
For a stationary state, $\partial_t \Psi_{2D}$ is given by $V_L/2\pi$, where $V_L$ is a constant corresponding to the externally applied tokamak loop voltage. 

As the difference between the 2D and the 3D $n=0$ state is not too large, we can linearize Equation~\eqref{induction4} around the corresponding 2D solution in order to extract 3D effects. This yields
\begin{align}
0 = -R \Delta \eta J_{\phi,2D}-R\eta_{2D}\Delta J_{\phi} +R \left[ \hat{\phi} \cdot \left( \mathbf{v}_1 \times \mathbf{B}_1 \right) \right]_{n=0} -R \left[ \eta_1 J_{\phi,1} \right]_{n=0}
\label{inductionLIN}
\end{align}
where the term $\partial_t \Delta \Psi$ has been dropped because it vanishes for stationary cases as well as for cases with a quasi-stationary periodic time evolution when it is time averaged over one period. Note, that this linearized induction equation is presented only to facilitate the understanding of the simulation results which result from fully 3D nonlinear calculations. 

The last term on the right of Equation~\eqref{inductionLIN} is negligible in the simulations as will be shown later. Out of the three remaining terms, the second term on the right describes the observed difference in the toroidal current density between a 3D and a 2D calculation and the two other terms represent mechanisms which can potentially be responsible for this difference. One possibility to obtain a flattening of the central current density profile is via a flattening of the central resistivity profile as described by the term that is proportional to $\Delta \eta$. In some of the simulations presented this is the leading effect as resistivity flattening is caused by a convective flattening of the temperature profile through the helical $(m=1,n=1)$ flow described above. This flow is also crucial for the second current flattening effect described by the third term on the right of Equation~\eqref{inductionLIN}. In this case the velocity perturbation combines with the perturbation of the magnetic field yielding a $n=0$ reduction of the background current density in the plasma center via a dynamo mechanism. This effect corresponds to an effective incremental negative loop voltage in the center of the tokamak opposing the externally applied loop voltage.

\begin{figure}
  \centering
  \includegraphics[width=0.49\textwidth]{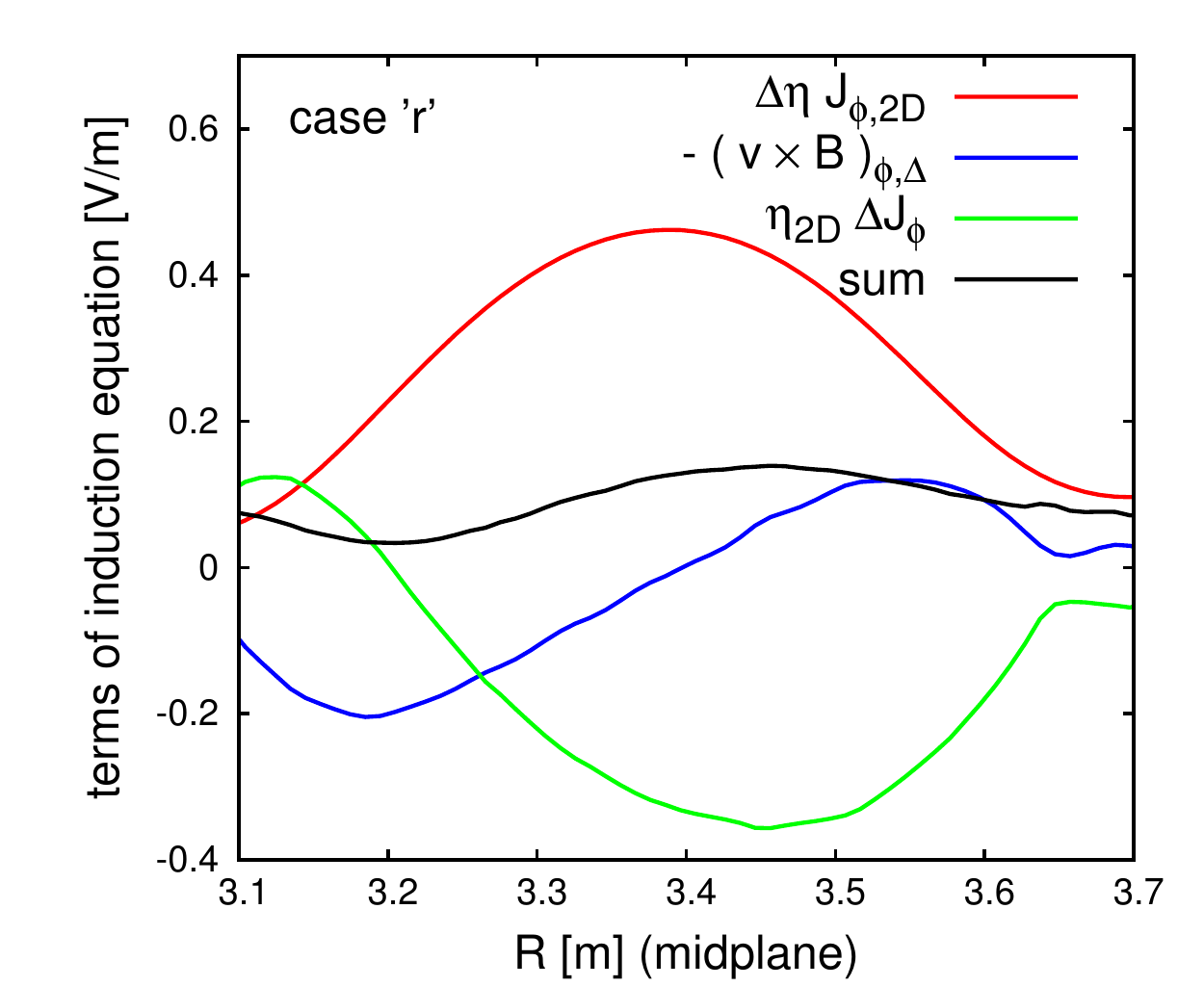}
  \includegraphics[width=0.49\textwidth]{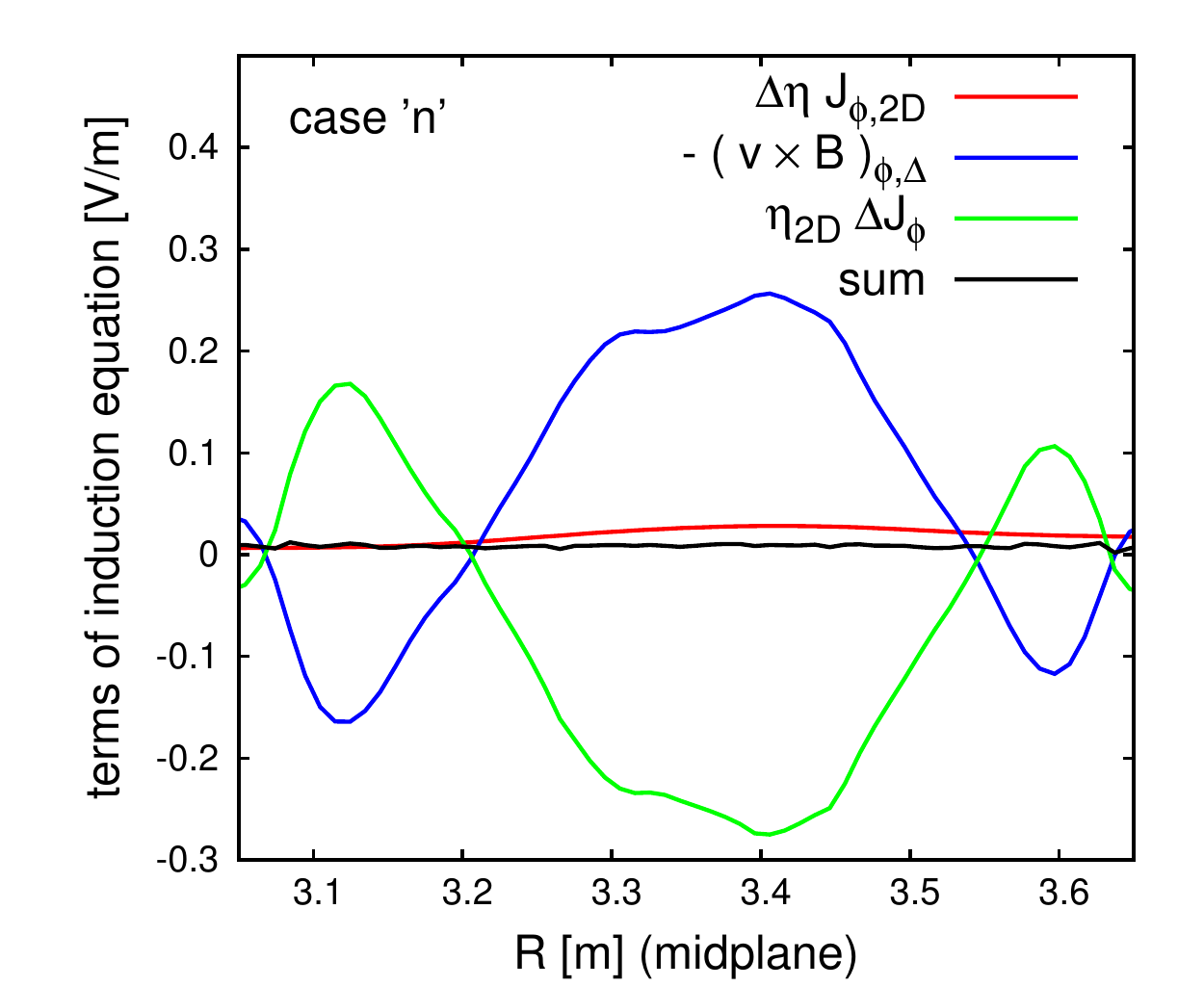}
  \caption{Terms of the linearized induction equation~\eqref{inductionLIN} in the plasma center in two simulations with different values of $\mu_0 \chi_\perp/\eta$. A flattening of the central $n=0$ current density profile can be caused by a flattening of the resistivity profile described by $\Delta \eta J_{\phi,2D}$ and by a dynamo loop voltage described by $-( \mathbf{v} \times \mathbf{B})_{\phi,\Delta}=-[ ( \mathbf{v} \times \mathbf{B})_{\phi,3D,n=0}-( \mathbf{v} \times \mathbf{B})_{\phi,2D}]$. Note that the sum of the three terms is negligible showing that these two effects fully account for the occurring change of the current density (i.e. the last term of Equation~\eqref{inductionLIN} is negligible). The case on the right has a higher value of $\chi_\perp$ and a proportionately stronger heat source yielding a stiffer temperature profile which decreases the effectiveness of convective resistivity flattening.}
  \label{termsinduction} 
\end{figure}

The strength of the different terms in Equation~\eqref{inductionLIN} in the plasma center is shown in Figure~\ref{termsinduction} for stationary states in two different simulations. In the case shown on the left the resistivity flattening effect is dominantly responsible for the diminished central toroidal current density, whereas in the other case the current flattening is predominantly caused by the dynamo loop voltage effect. The difference between these two cases is the value of the perpendicular heat diffusion coefficient $\chi_{\perp}$ and correspondingly the strength of the heat source. They are higher in the second case which leads to a stiffer temperature profile that cannot easily be altered by convection. Therefore, the resistivity flattening effect does not play a significant role in the second case. The temperature gradient remains large, driving the instability  that results in the dynamo driven loop voltage. In modern large tokamaks, and in particular in Hybrid discharges which are characterized by strong heating and large turbulent heat flux, the ratio of the resistive time scale to the heat diffusion time scale $\mu_0 \chi_\perp / \eta$ is in a regime where the dynamo effect is dominant as in the case on the right. 

\section{Conditions for flux pumping}
\label{conditions}
In order to analyze under which conditions the flux pumping mechanism described above is sufficiently strong to prevent sawtoothing, we present a set of 3D nonlinear MHD simulations run to their asymptotic states. Three parameters have been varied: 
(1) The poloidal beta defined as
\begin{align}
\beta_{p1} = \frac{2\mu_0}{B^2_{\theta}} \int_{0}^{r_1} \left( \frac{r}{r_1} \right)^2 \left(  -\td{p}{r} \right) \, dr
\end{align}
where $r$ is the midplane minor radius and $r_1$ is the radius where the velocity perturbation vanishes and $q$ first differs from unity. The value of $\beta_{p1}$ determines the drive of the instability that enables flux pumping by generating the necessary helical flow. 
(2) The ratio of the resistive time scale to the heat diffusion time scale $\mu_0 \chi_\perp / \eta$ has been varied by varying $\chi_\perp$ at the same rate as the strength of the applied heat source while keeping $\eta$ fixed. As mentioned before, this quantity controls the stiffness of the temperature profile and thus the effectiveness of the resistivity flattening effect. 
(3) The third parameter varied is the peakedness of the heat source. In a 2D simulation this parameter determines the value of $q_0$ which is a measure for how strong the current flattening mechanism in a 3D simulation needs to be in order to keep $q_0=1$. We define $\Delta_{2D}$ as the corresponding rate of magnetic flux pumping:
\begin{align}
\Delta_{2D} &= -\frac{2\eta B_{\phi,\tn{axis}}}{\mu_0 R_{\tn{axis}}} \frac{\left( 1-q_{0,2D} \right)}{q_{0,2D}} \nonumber \\ &\approx \eta J_{\phi,\tn{axis}}(q_0=1)-\eta J_{\phi,\tn{axis}}(q_0=q_{0,2D}) \, .
\end{align}
The different types of asymptotic behavior resulting from our simulations are discussed in the following.

\begin{figure}
  \centering
  \includegraphics[width=0.49\textwidth]{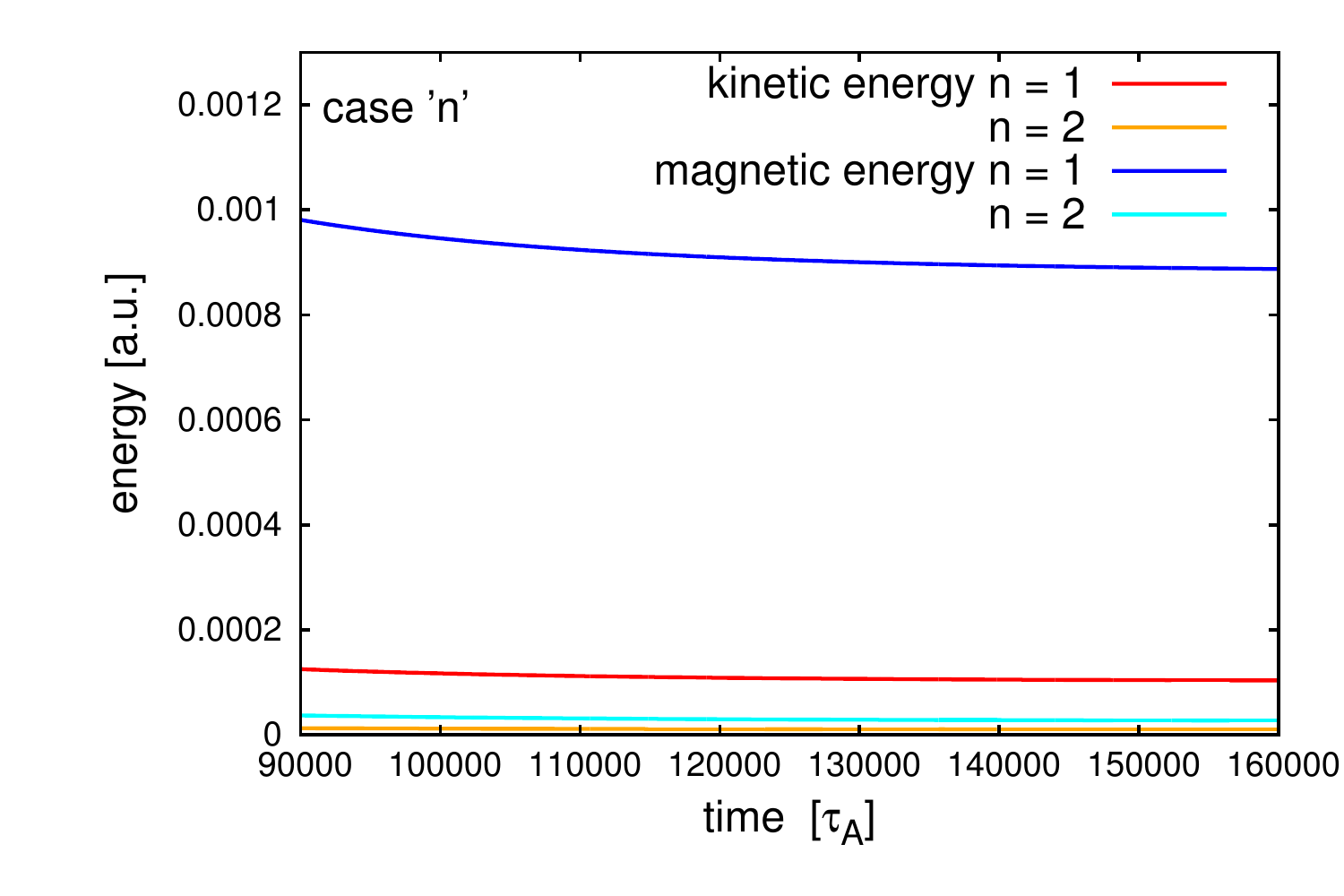}
  \includegraphics[width=0.49\textwidth]{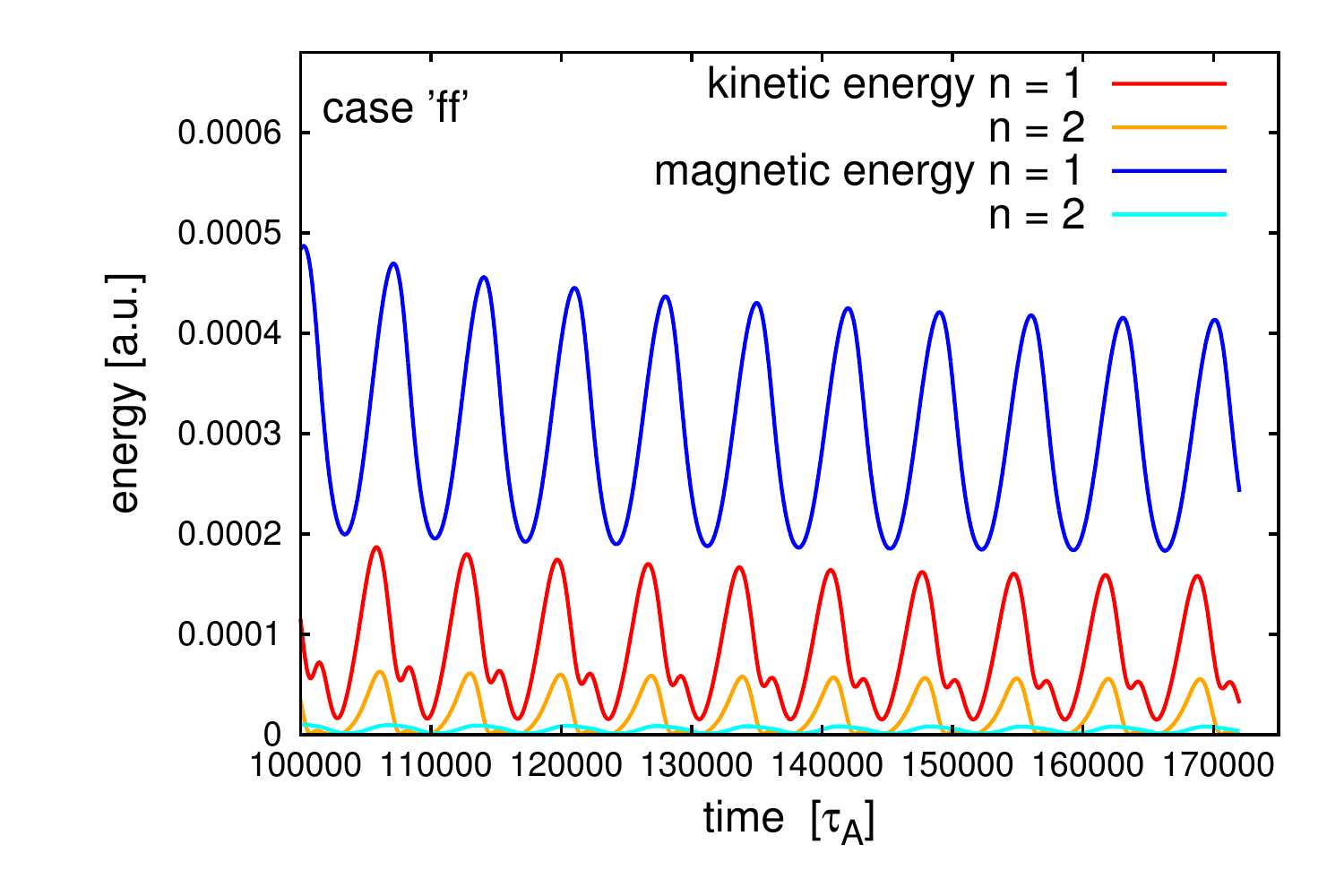}
  \caption{Kinetic and magnetic energies of the $n=1$ and $n=2$ harmonics for a stationary case (left) and for a quasi-stationary oscillatory case (right). In both cases sawtoothing is prevented.}
  \label{energies} 
\end{figure}

\begin{figure}
  \centering
  \includegraphics[width=0.49\textwidth]{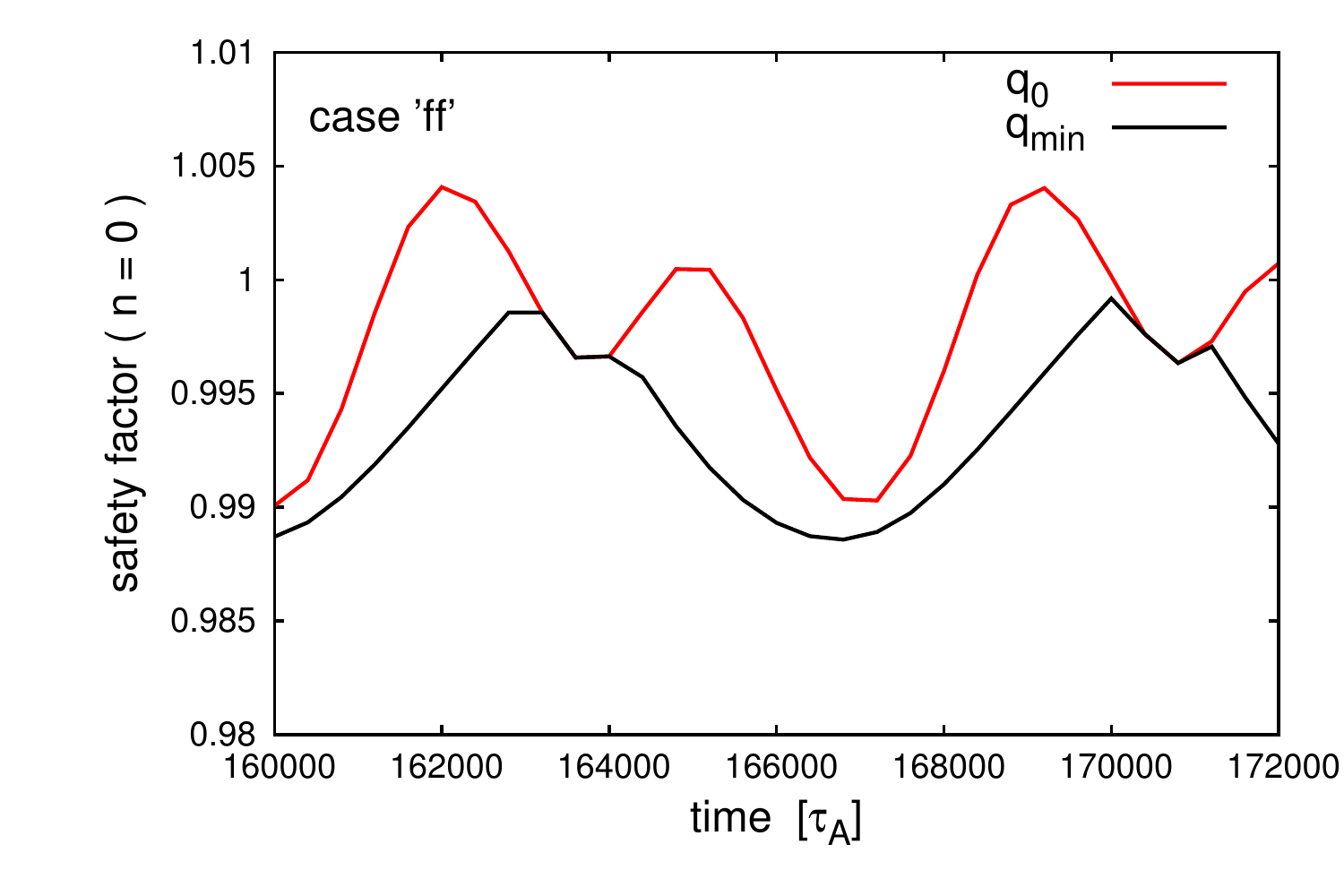}
  \includegraphics[width=0.49\textwidth]{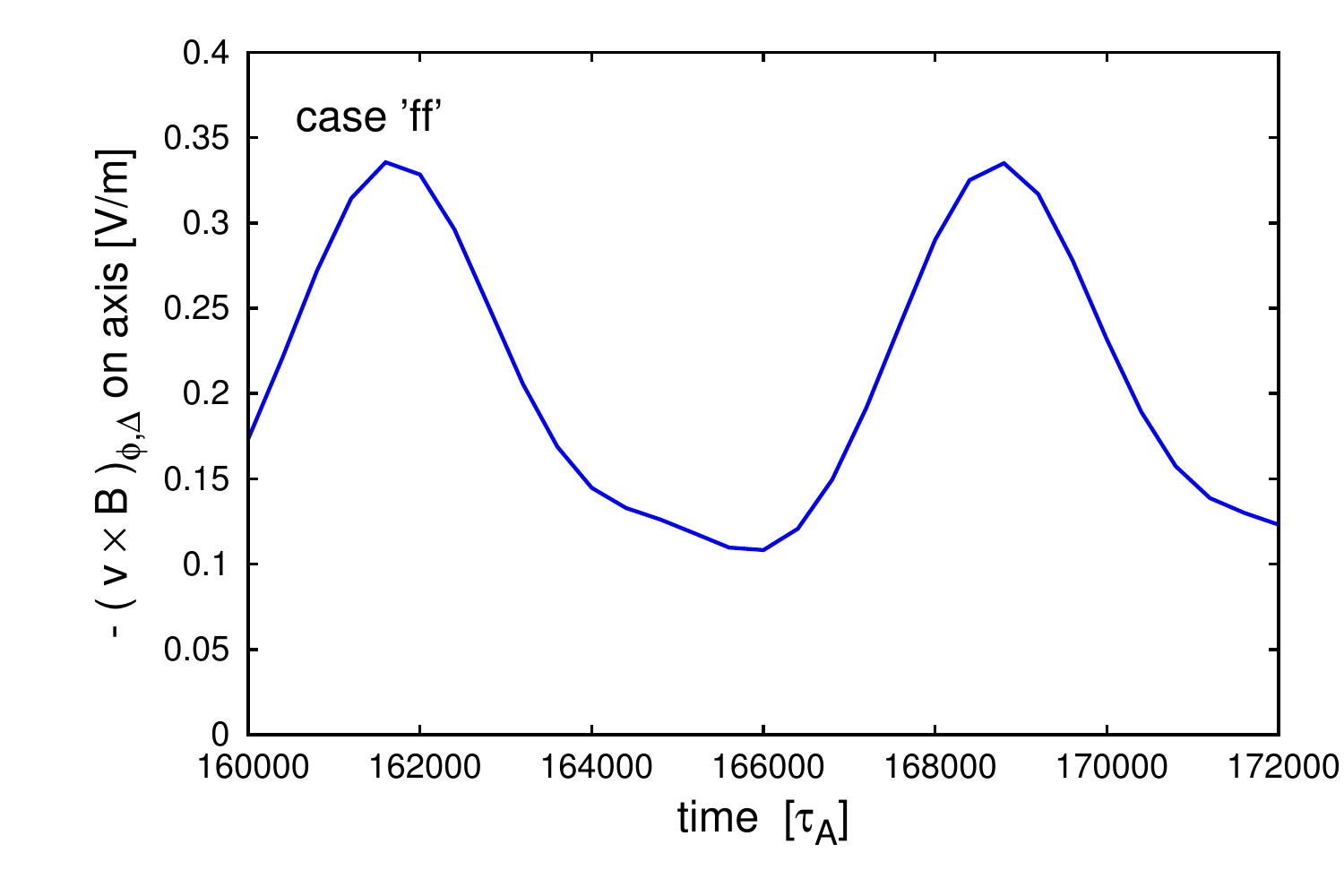}
  \caption{Time evolution of the value of $q$ on axis and the minimum value of $q$ (left) and of the strength of the dynamo loop voltage term on axis (right) for a quasi-stationary case. Despite the oscillation, the dynamo loop voltage effect is strong enough to keep the safety factor profile in the central plasma region flat with values close to unity at all times.}
  \label{osc} 
\end{figure}

\textbf{\emph{Sawtooth-free states}}. 
One possibility for an asymptotic state is the sawtooth-free time-independent state with a flat central safety factor close to unity characterized by a $(m=1,n=1)$ convection cell in the plasma center as described in Section~\ref{mechanism}. We also find a slight variation of this behavior which features the same characteristics, but superimposed with an oscillation. The magnetic and kinetic energies in the first two toroidal harmonics for a stationary and an oscillatory case are shown in Figure~\ref{energies}. The strength of the dynamo loop voltage effect as well as the time evolution of the minimum value of the safety factor ($q_{\tn{min}}$) and $q_0$ for an oscillatory case are shown in Figure~\ref{osc}. It can be seen that despite the oscillation, the central safety factor profile is still very close to unity at all times so that sawtoothing is prevented. Note that in contrast to sawtooth oscillations, in these cases the magnetic field line structure is not rearranged by a reconnection process.

\begin{figure}
  \centering
  \includegraphics[width=0.49\textwidth]{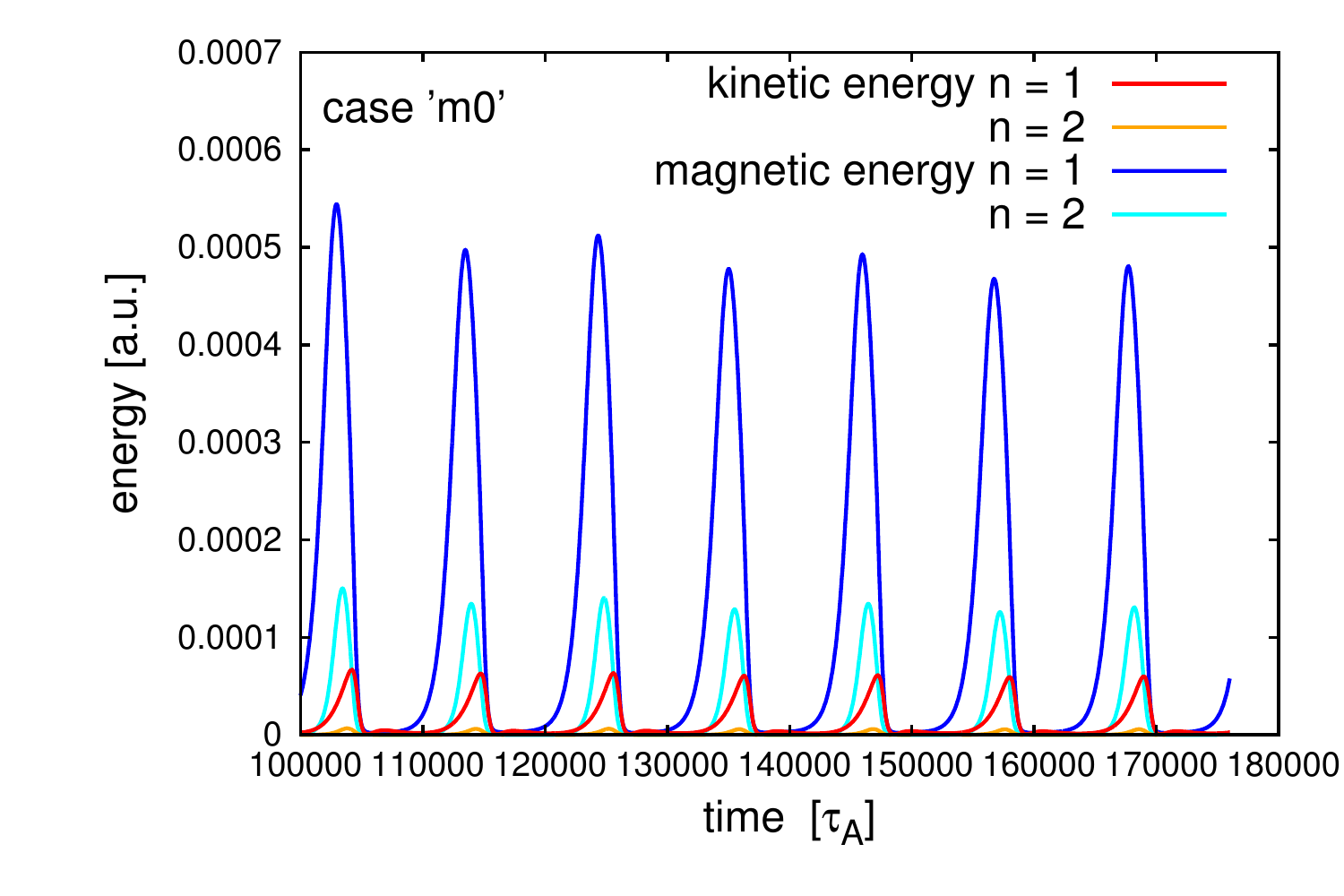}
  \includegraphics[width=0.49\textwidth]{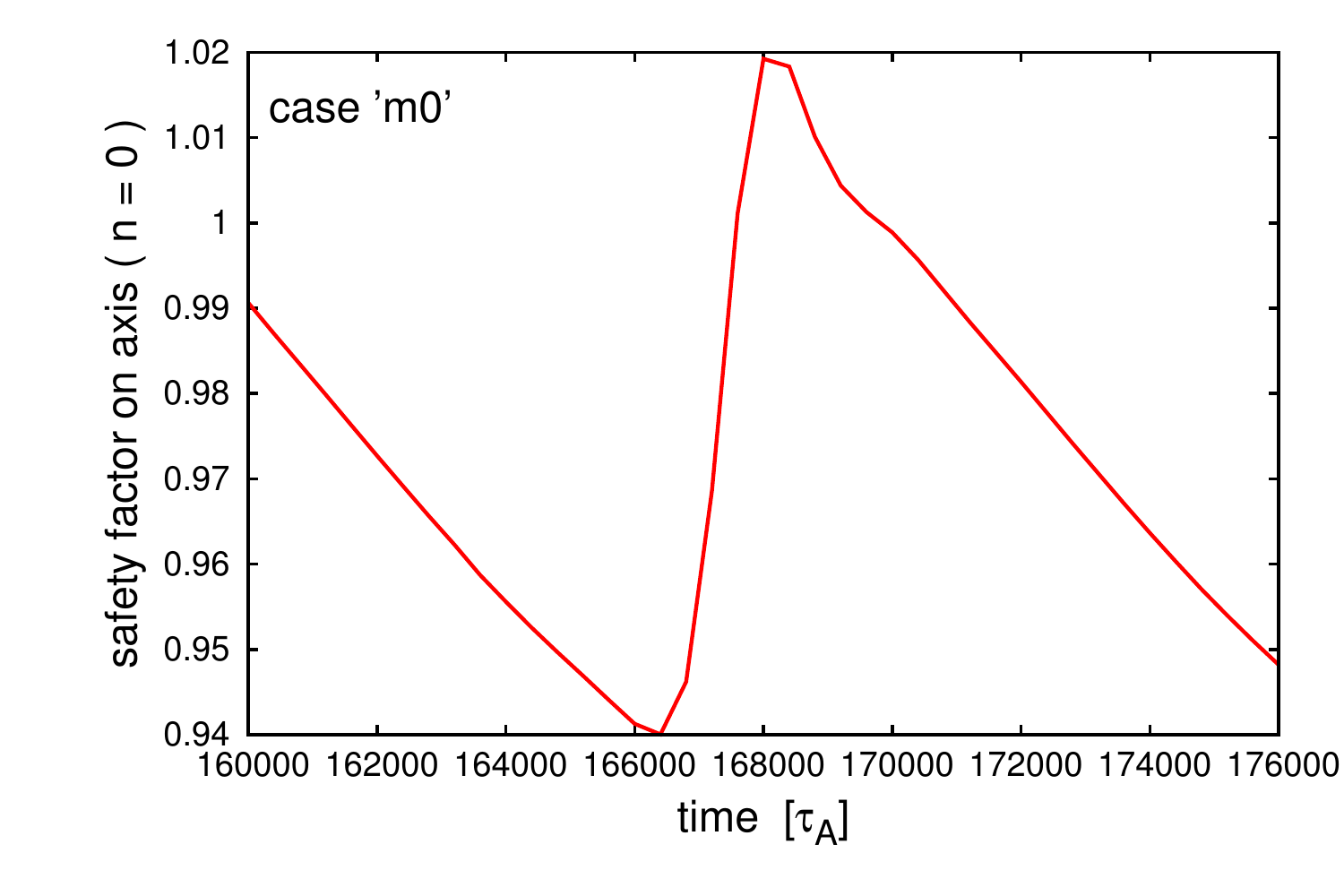}  
  \caption{Kinetic and magnetic energies of the $n=1$ and $n=2$ harmonics (left) and time evolution of $q_0$ during about one cycle (right) for a case exhibiting a sawtooth-like behavior. After each crash $q_0 \approx 1$ and axisymmetry is restored.}
  \label{cycling} 
\end{figure}

\begin{figure}
  \centering
  \includegraphics[height=0.34\textwidth,trim={2.9cm 0 3.5cm 0},clip]{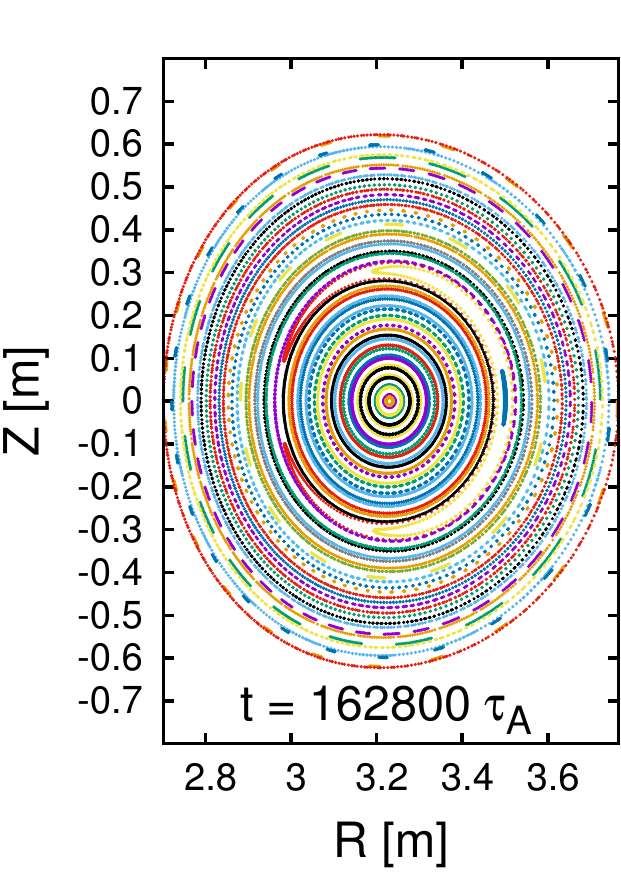}
  \includegraphics[height=0.34\textwidth,trim={3.9cm 0 4.08cm 0},clip]{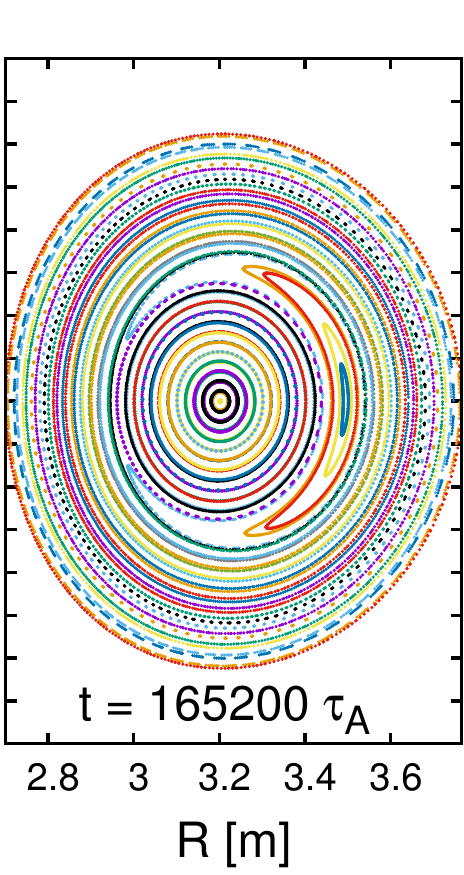}
  \includegraphics[height=0.34\textwidth,trim={3.9cm 0 4.08cm 0},clip]{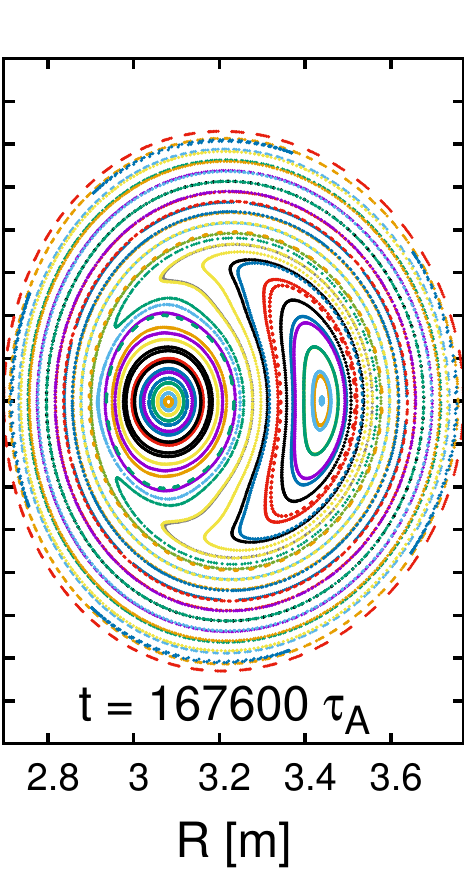}
  \includegraphics[height=0.34\textwidth,trim={3.9cm 0 4.08cm 0},clip]{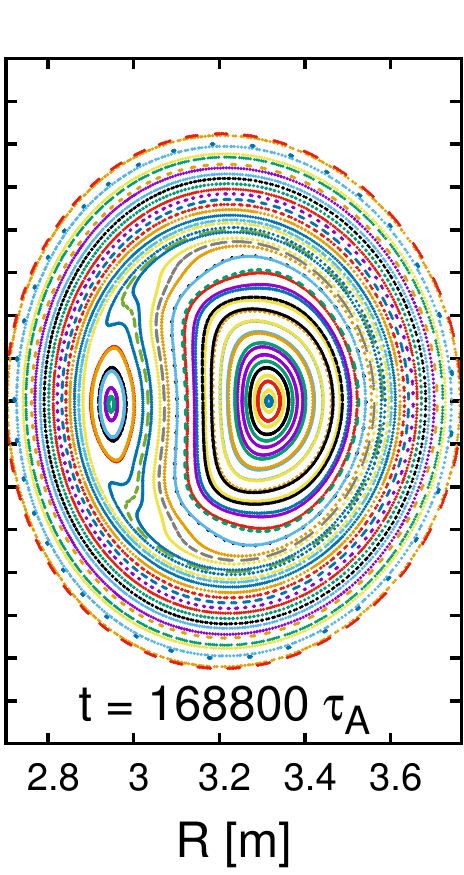}
  \includegraphics[height=0.34\textwidth,trim={3.9cm 0 4.08cm 0},clip]{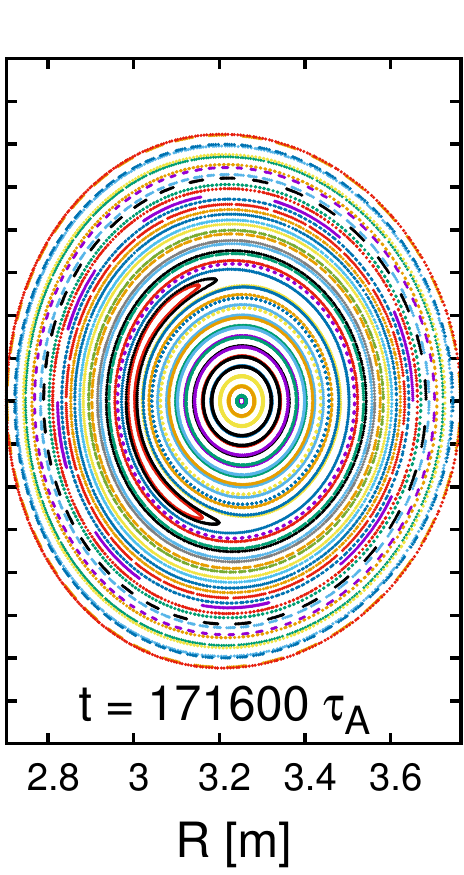}
  \caption{Poincar\'e plots showing the magnetic field line structure in the central plasma region at different points in time during a sawtooth cycle. As described in Kadomtsev's model, the $(m=1,n=1)$ magnetic island grows until it has entirely replaced the original plasma core. (Case \lq m0\rq).}
  \label{poincarekadomtsev} 
\end{figure}

\textbf{\emph{Sawtooth-like behavior}}.
This is different for the sawtoothing cases where the safety factor on axis decreases to values significantly below unity such that $q_0\approx 1$ is restored in periodically repeating crashes. The time traces of the $n=1$ magnetic and kinetic energies as well as $q_0$ for such a case are shown in Figure~\ref{cycling}. The evolution of the magnetic topology is in agreement with Kadomtsev's full reconnection model as can be seen from the Poincar\'e plots for different points in time during one cycle in Figure~\ref{poincarekadomtsev}. As the Lundquist number used is significantly below its realistic value and two-fluid effects are not included into our calculations, several characteristics of realistic sawtooth crashes like the fast crash times as found in \cite{Halpern2011,Guenter2014,Yu2015} are not expected to be reproduced in these simulations.

\begin{figure}
  \centering
  \includegraphics[height=0.32\textwidth,trim={2.8cm 0 3.3cm 0},clip]{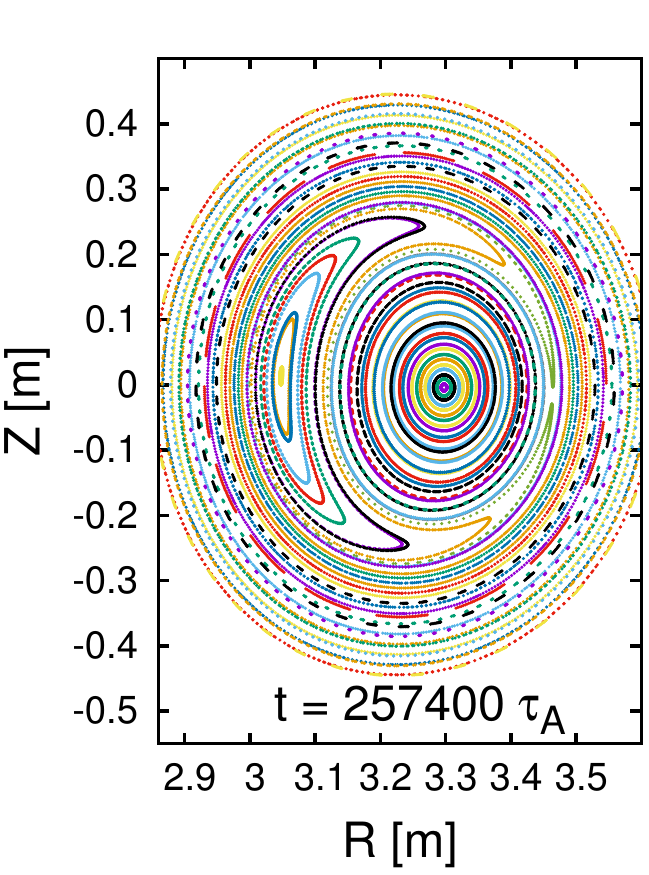}
  \includegraphics[height=0.32\textwidth,trim={3.8cm 0 3.9cm 0},clip]{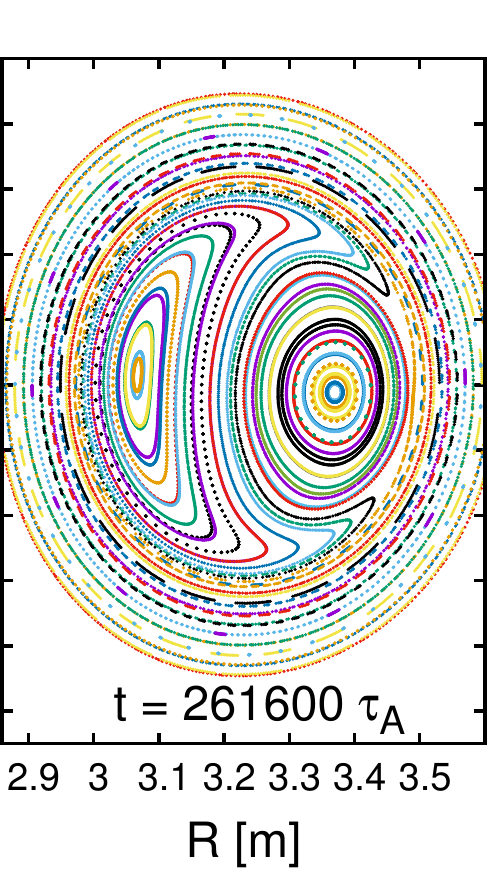}
  \includegraphics[height=0.32\textwidth,trim={3.8cm 0 3.9cm 0},clip]{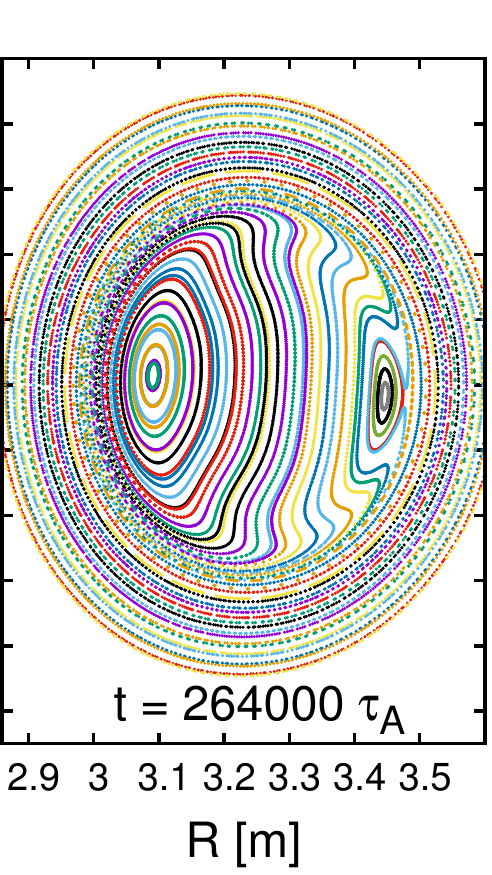}
  \includegraphics[height=0.32\textwidth,trim={3.8cm 0 3.9cm 0},clip]{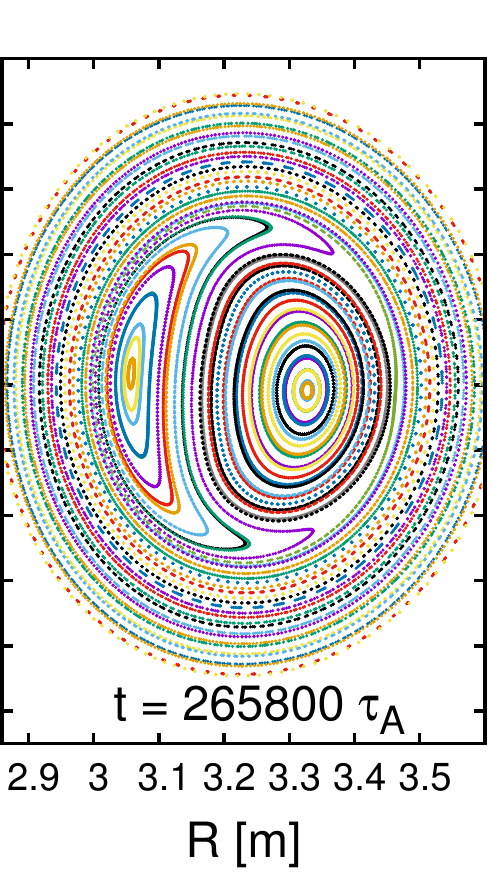}
  \includegraphics[height=0.32\textwidth,trim={3.8cm 0 3.9cm 0},clip]{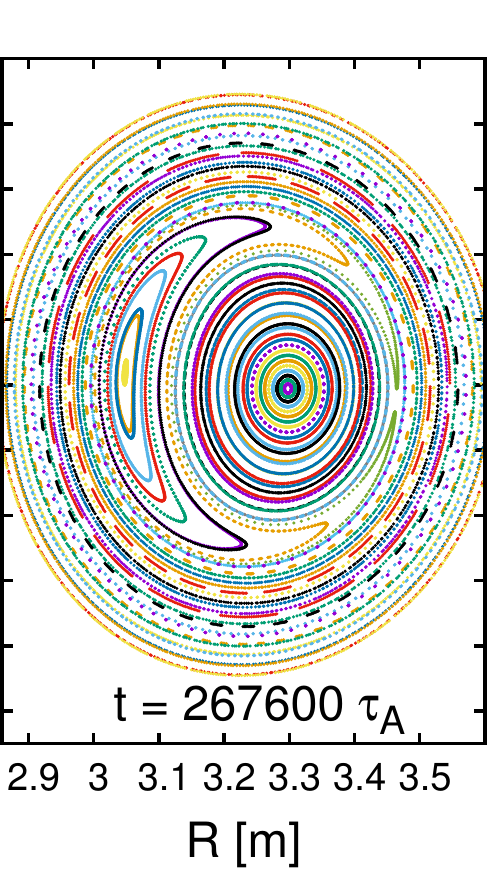}
  \caption{Poincar\'e plots showing the magnetic field line structure in the central plasma region at different points in time during an incomplete sawtooth cycle. Before the $(m=1,n=1)$ magnetic island can replace the original plasma core, the reconnection process stops and reverses. Note, that this series of plots cover one entire cycle showing that an axisymmetric state is not reached at any time. (Case \lq m3\rq).}
  \label{poincareincomplete} 
\end{figure}

\begin{figure}
  \centering
  \includegraphics[width=0.49\textwidth]{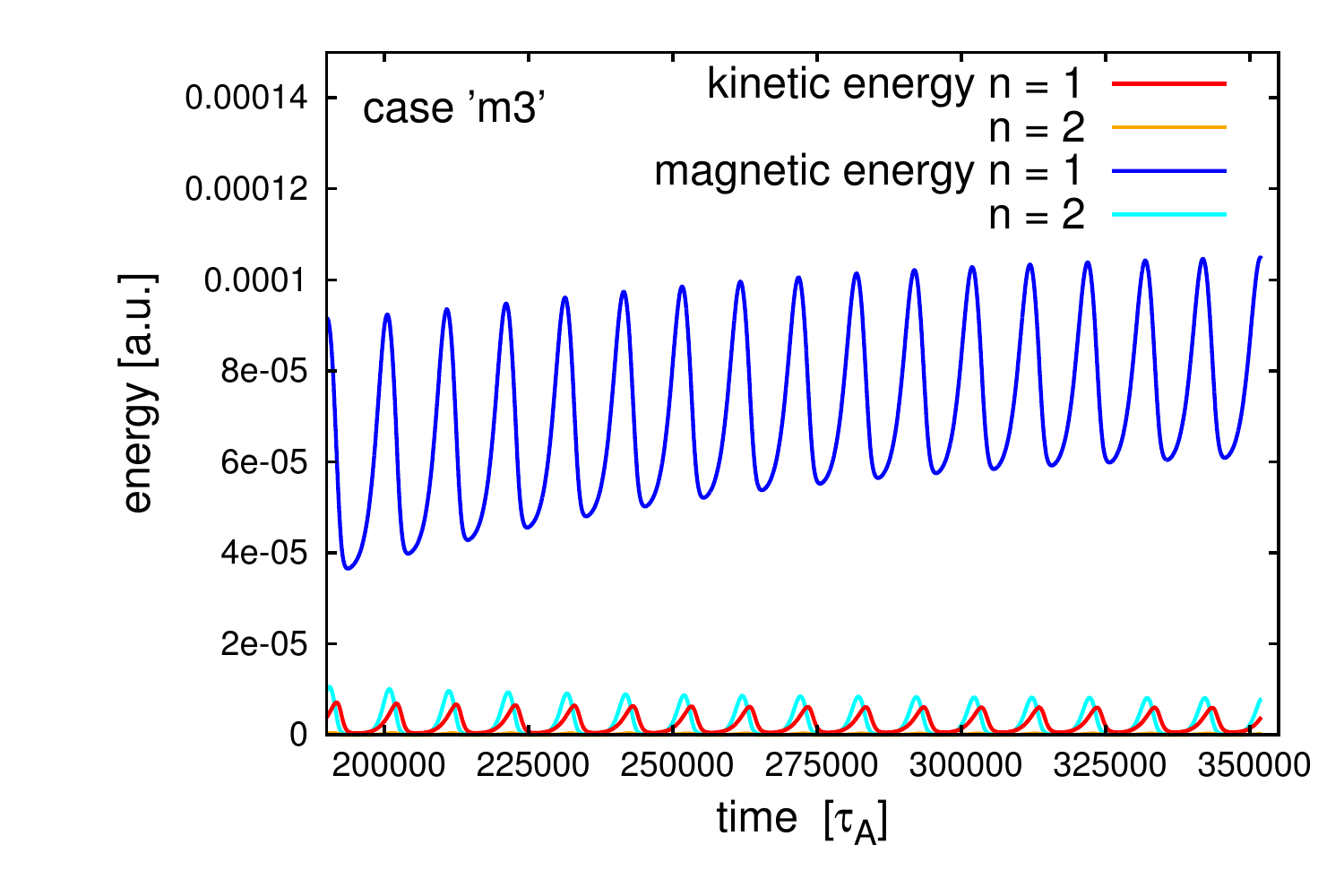}
  \includegraphics[width=0.49\textwidth]{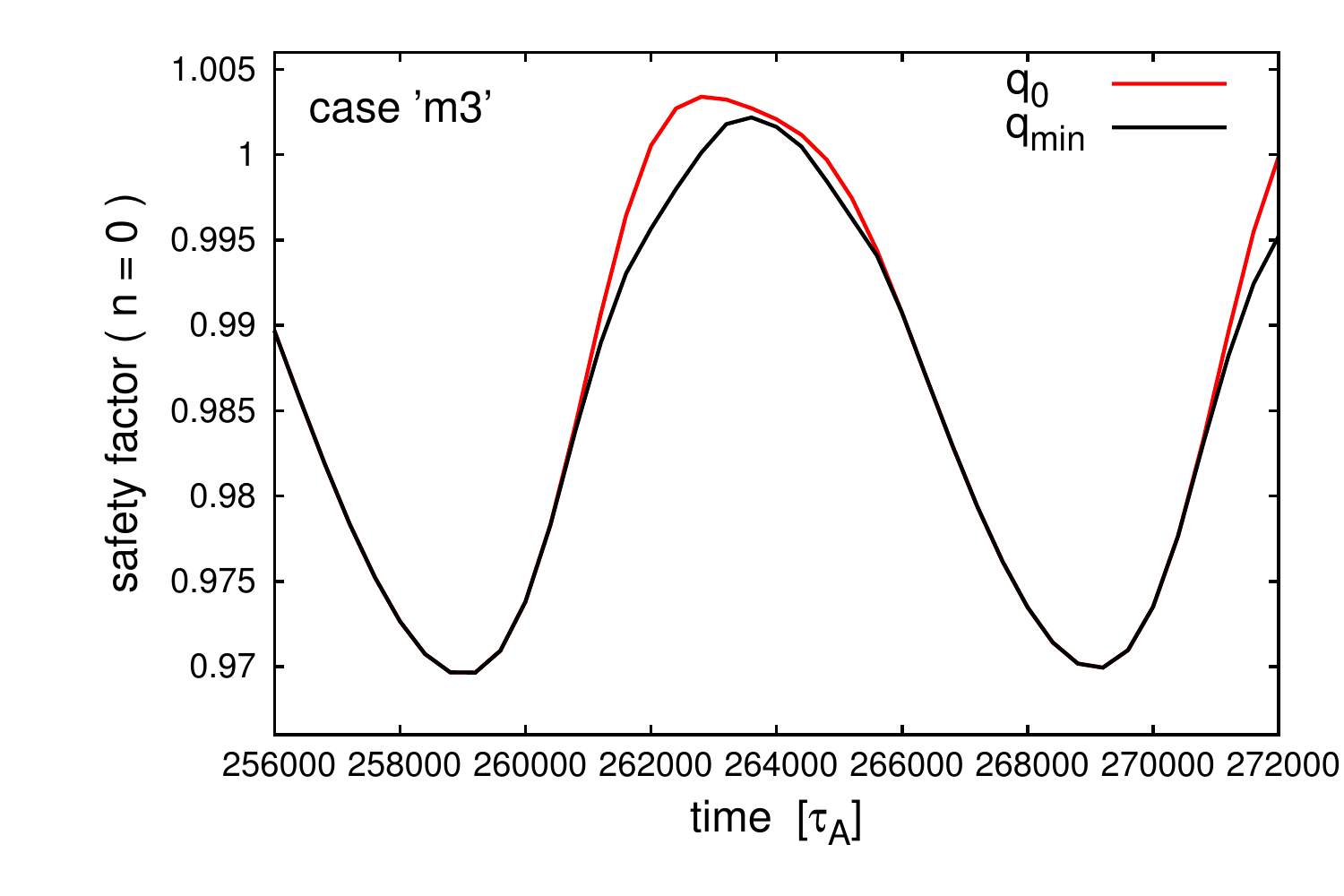}  
  \caption{Kinetic and magnetic energies of the $n=1$ and $n=2$ harmonics (left) and time evolution of $q_0$ and $q_{\tn{min}}$ during about two cycles (right) for a case exhibiting incomplete sawtooth reconnection. Like in sawtoothing cases as \lq m0\rq, $q_0$ cannot be prevented from dropping to values significantly below unity leading to the growth of the $(m=1,n=1)$ magnetic island. However, since the reconnection process does not complete, there is a finite offset in the $n=1$ magnetic energy.}
  \label{incomplete} 
\end{figure}

\begin{figure}
  \centering
  \includegraphics[width=0.49\textwidth]{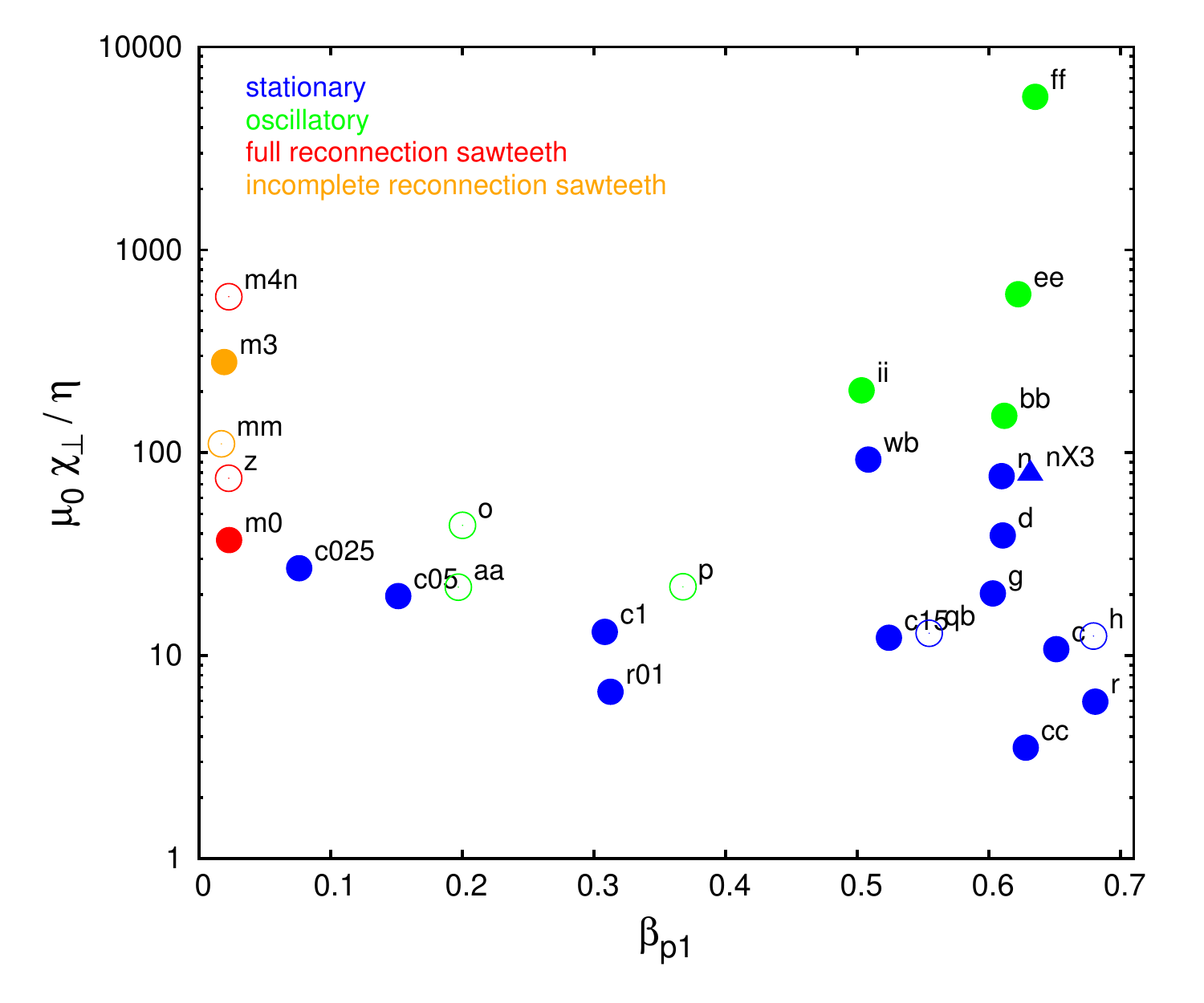}
  \includegraphics[width=0.49\textwidth]{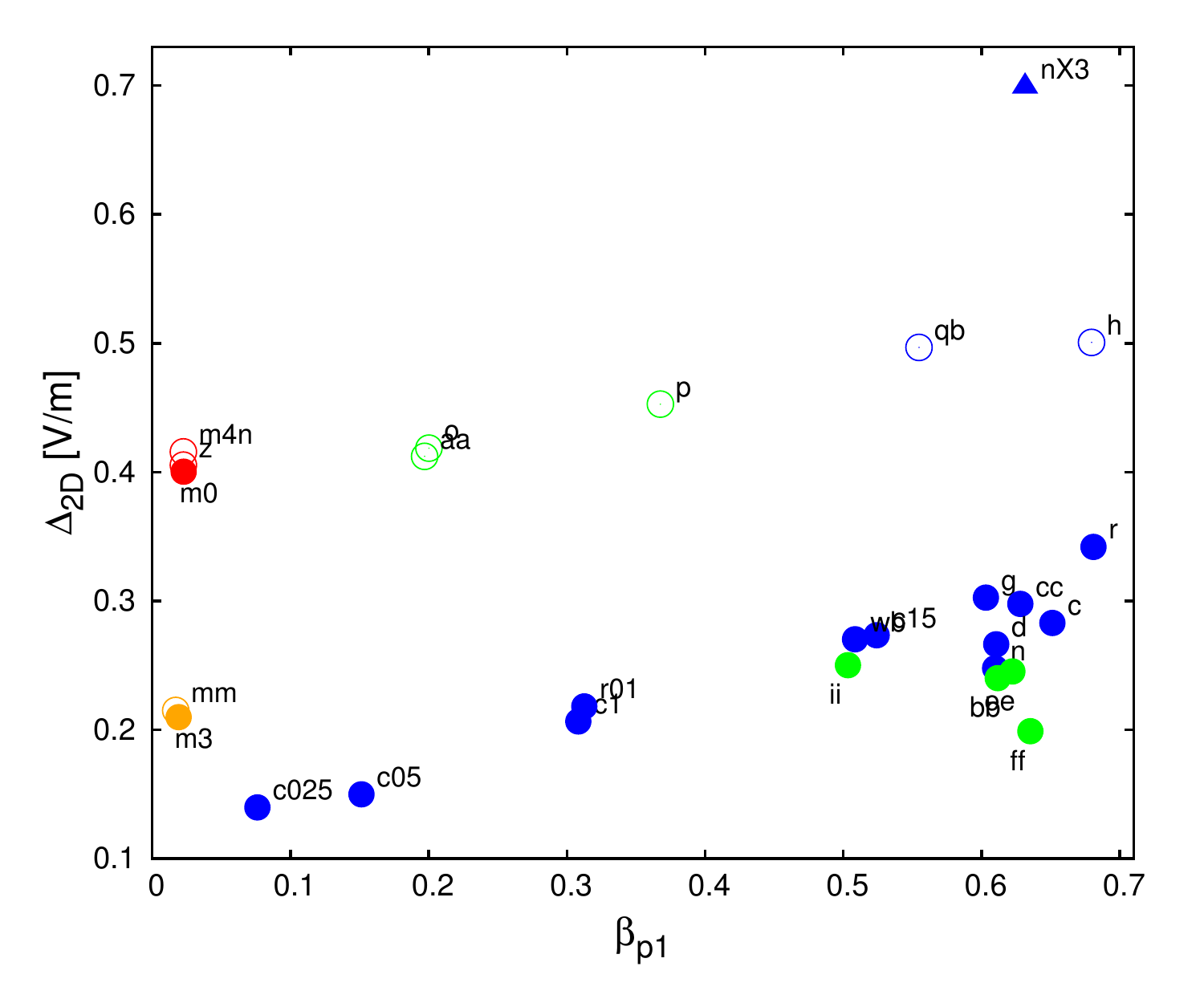}
  \caption{Overview of the covered parameter space. Each data point corresponds to a 3D nonlinear simulation run to its asymptotic state. Stationary and quasi-stationary sawtooth-free cases are marked in blue and green, respectively. Cases exhibiting complete and incomplete sawtooth reconnection are marked in red and orange, respectively. Open symbols correspond to cases with a more peaked heat source profile. The case marked by a triangle has a three times larger resistivity. Note, that for cases with $\beta_{p1} <0.03$ and $\mu_0 \chi_\perp /\eta <1.5 \cdot 10^2$, Ohmic heating plays a role in determining the heating profile.}
  \label{beta-kappa_Delta} 
\end{figure}

In another type of sawtoothing cases that we find, the sawtooth cycle starts similarly as described above with a decrease of $q_0$ destabilizing an internal kink, but instead of completing, the reconnection process stops and reverses. The corresponding evolution of the magnetic topology is shown in Figure~\ref{poincareincomplete}. In these cases axisymmetry is not recovered after the crashes which manifests itself in an offset in the $n=1$ magnetic energy as shown in Figure~\ref{incomplete}. As can be seen from the second plot in Figure~\ref{beta-kappa_Delta}, these incomplete sawtooth reconnection cases only occur at low values of $\Delta_{2D}$ which corresponds to low linear drive for the internal kink instability. It is possible that in a more complete model than the one used for the presented calculations, a similar behavior occurs also at higher $\Delta_{2D}$ if the internal kink is stabilized by a more realistic physics model including, for example, diamagnetic drift, finite Larmor radius or energetic particle effects \cite{Migliuolo1993}. It is to be investigated in future work if this might then provide a possible explanation for the experimental observations indicating incomplete sawtooth reconnection.

\begin{figure}
  \centering
  \includegraphics[width=0.7\textwidth]{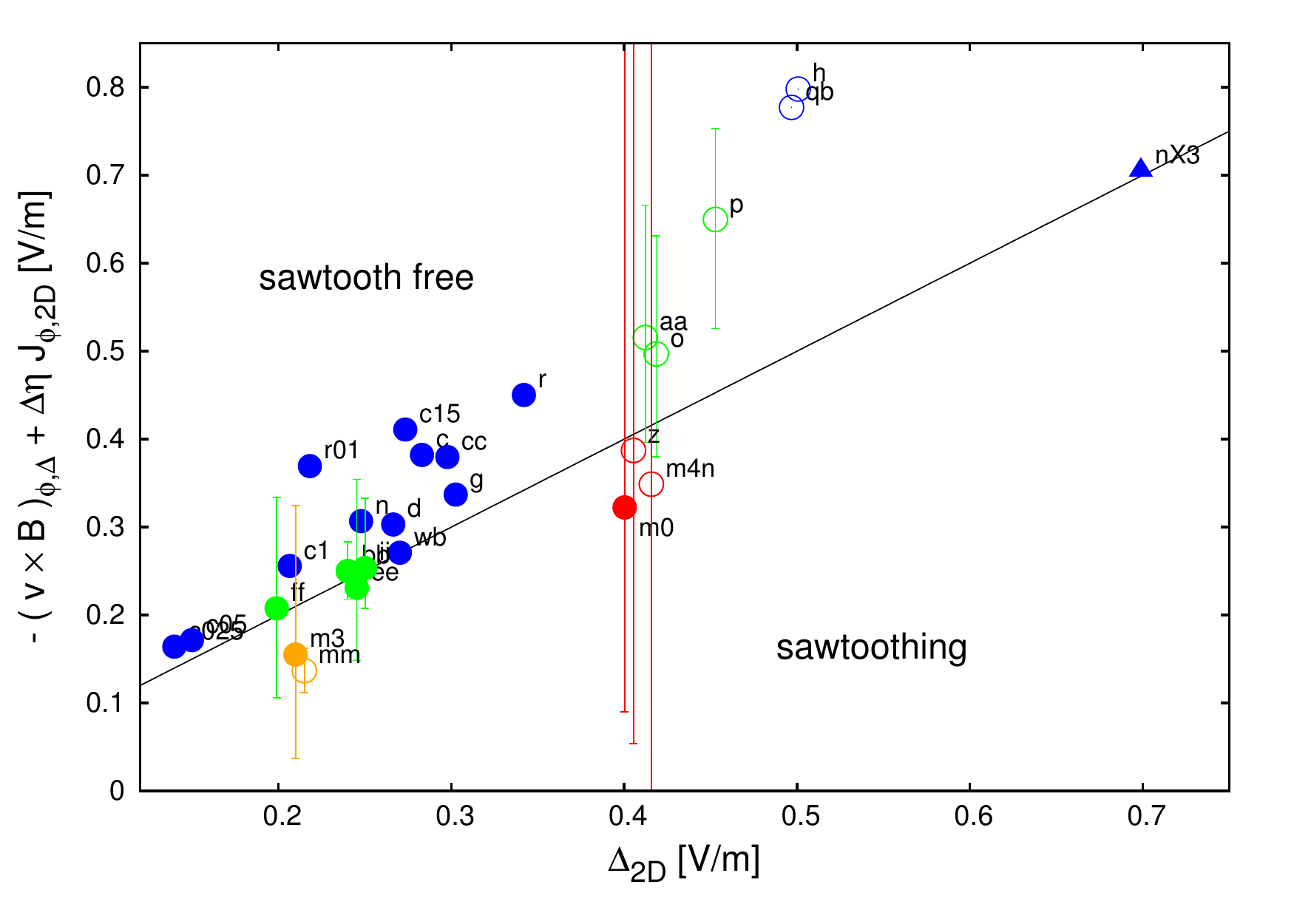}
  \caption{Combined strength of magnetic flux pumping effects on axis versus the amount of flux pumping which is necessary to keep $q_0=1$ for the different 3D nonlinear simulations. The black line indicates where the two quantities are equal. The sawtooth-free cases (blue and green) lie at or above this threshold, whereas the sawtoothing cases (red and orange) are found below. As the strength of the flux pumping effects varies with time for the oscillating and the sawtoothing cases, error bars indicate the range of their oscillation and the data points are set to their time-average over one period.}
  \label{Delta-VXBDeta} 
\end{figure}

\begin{figure}
  \centering
  \includegraphics[width=0.45\textwidth]{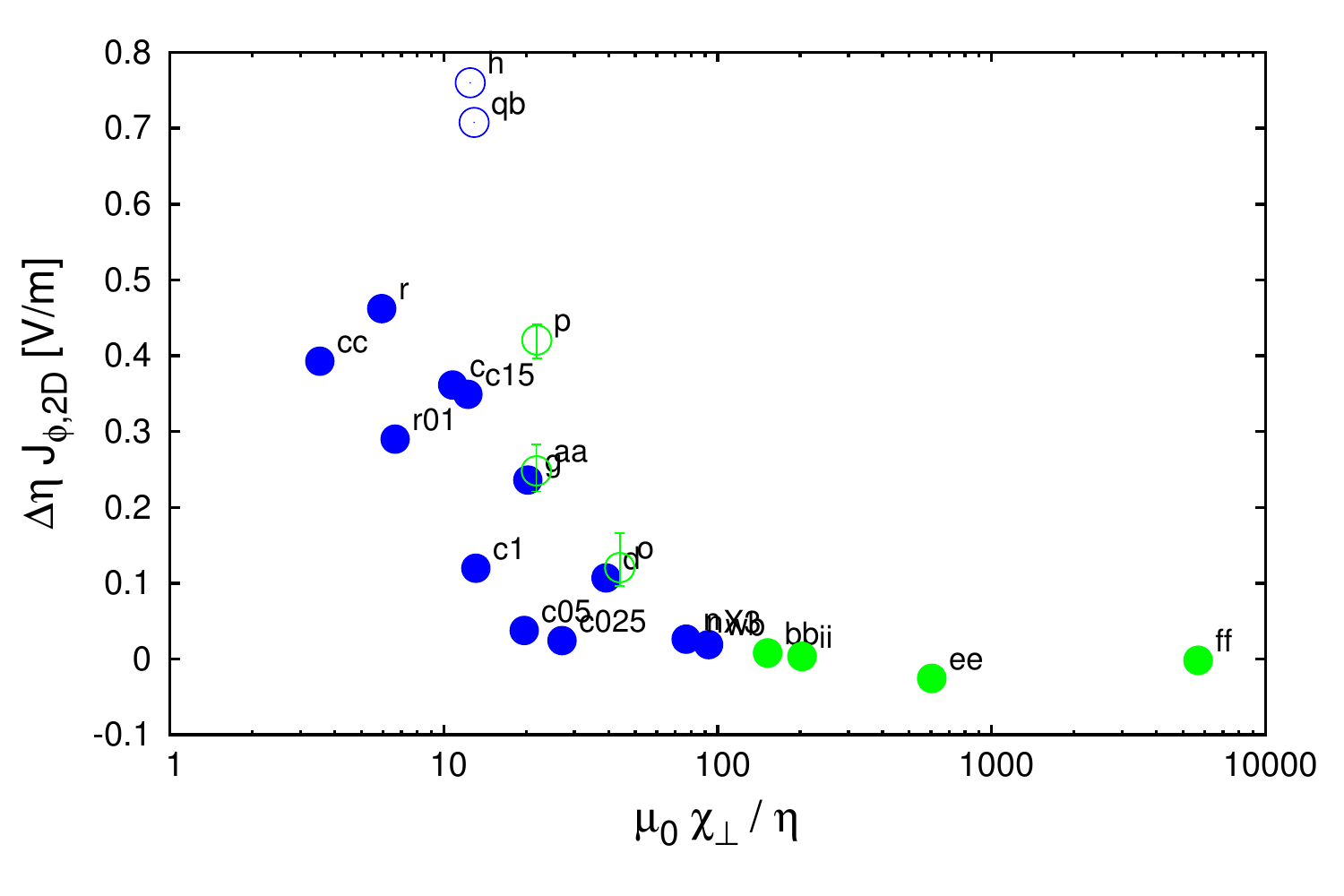}
  \includegraphics[width=0.45\textwidth]{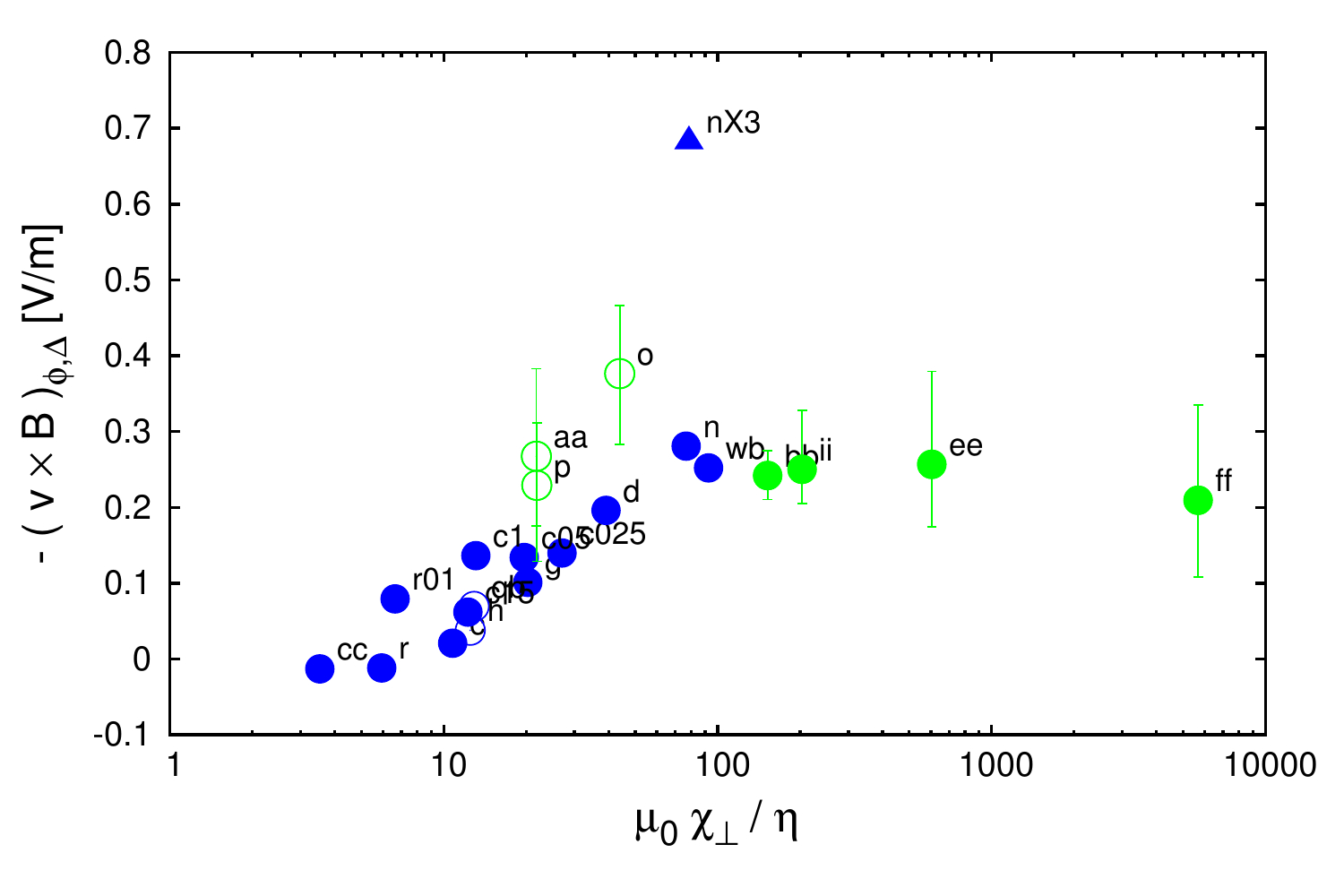}
  \caption{Strength of the resistivity flattening effect (left) and the dynamo loop voltage (right) on axis for different sawtooth-free cases. The value of $\chi_\perp$ which is varied together with the strength of the heat source controls the stiffness of the temperature profile.}
  \label{kappa-VXB_Deta} 
\end{figure}

We now focus on analyzing under which conditions the flux pumping mechanisms are strong enough to be able to prevent sawtoothing in the simulations. In Figures~\ref{beta-kappa_Delta}-\ref{kappa-VXB_Deta}, each data point corresponds to one 3D nonlinear MHD simulation. According to both plots in Figure~\ref{beta-kappa_Delta}, sawtooth-free states only occur at sufficiently high $\beta_{p1}$. Below that threshold in $\beta_{p1}$, the pressure-driven instability at low magnetic shear is not strong enough to provide the helical flow necessary for the flux pumping mechanisms to work. The existence of a threshold in $\beta$ is consistent both with the simulation results presented in \cite{Halpern2010} and with Hybrid discharges being characterized by high values of $\beta$.  

As illustrated in Figure~\ref{Delta-VXBDeta}, we find that sawteeth are avoided only if the combined strength of the two current flattening mechanisms equals or exceeds the rate of magnetic flux change needed to keep $q_0\approx 1$, a quantity which is approximately given by $\Delta_{2D}$. For quasi-stationary cases, the strength of the current flattening terms has been time-averaged over one period. 

In Figure~\ref{kappa-VXB_Deta} the two mechanisms are separated showing that for increasing $\mu_0 \chi_\perp/\eta$ the strength of the resistivity flattening effect decreases whereas the dynamo loop voltage effect strengthens. As discussed before, the former trend is due to the decreased effectiveness of the convective flattening of the temperature profile for high values of $\chi_\perp$ and strong heat sources. The decrease of the strength of the dynamo voltage effect for low $\chi_\perp$ is due to the fact that the strong resistivity flattening effect already provides enough current flattening to keep $q_0\approx 1$. An additional dynamo loop voltage would further increase the central value of the safety factor above unity. This would stabilize the low-shear pressure-driven instability which needs $q\approx 1$ and thus weaken the helical flow responsible for the dynamo voltage. In this way the strength of the dynamo loop voltage effect is self-regulated to always provide a flat safety factor profile close to unity in the plasma center. Note that in Figure~\ref{kappa-VXB_Deta}, only cases with a similar value of $\Delta_{2D}$ should be compared to each other as the necessary amount of flux pumping depends on this parameter. 

\section{Linear analysis}
\label{linear}

In the previous sections it has been discussed how in 3D nonlinear MHD simulations a saturated quasi-interchange instability allowed for by low central magnetic shear and $q_0\approx 1$, gives rise to perturbations of the velocity and the magnetic field which can generate an effective loop voltage via a dynamo effect. By means of a linear stability analysis of an equilibrium featuring such a safety factor profile, it is confirmed that the most unstable mode can be characterized as a quasi-interchange instability and that the linear perturbations of the velocity and magnetic field can be combined to calculate a corresponding dynamo loop voltage term.
\begin{figure}
  \centering
  \includegraphics[width=0.48\textwidth]{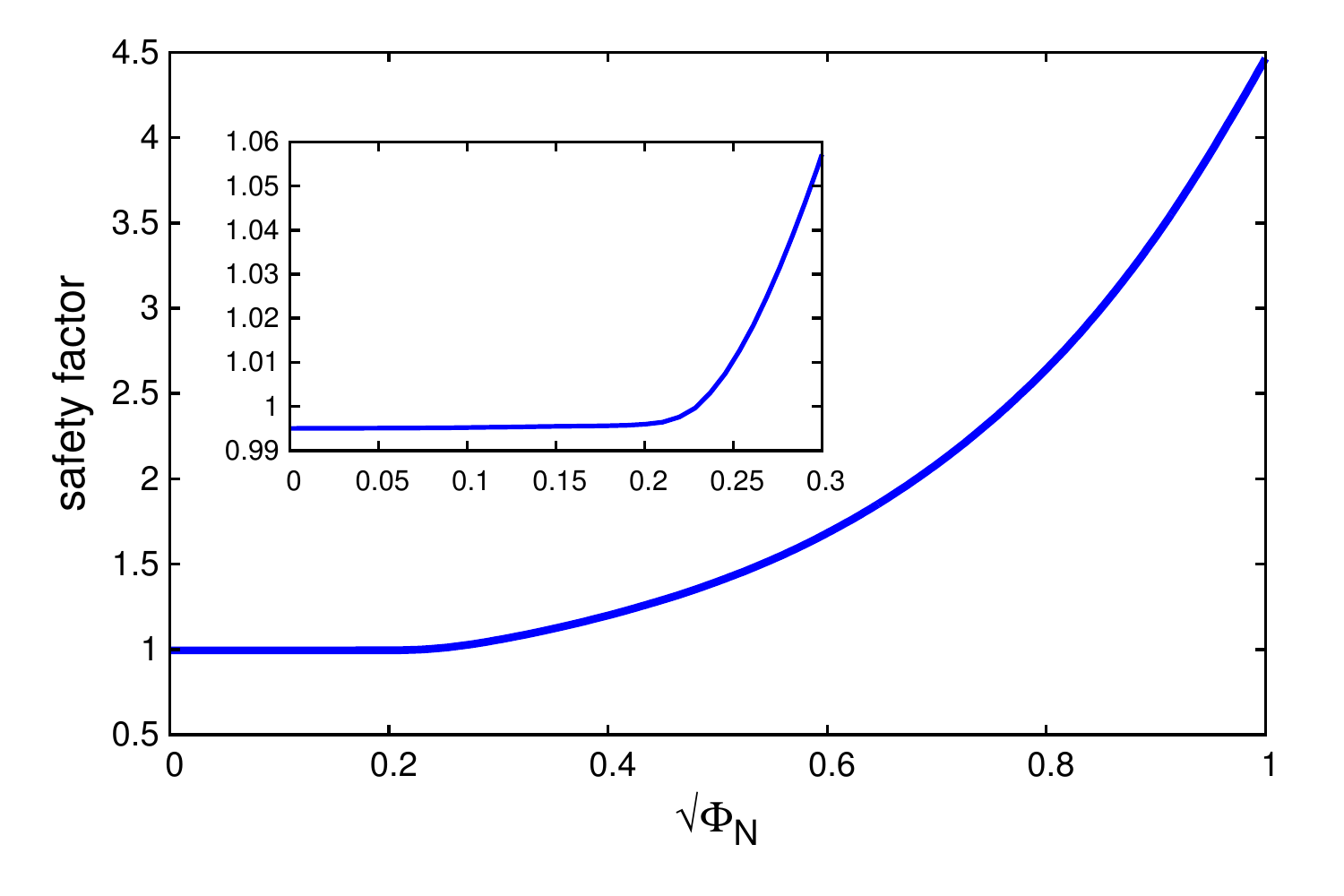}
  \includegraphics[width=0.48\textwidth]{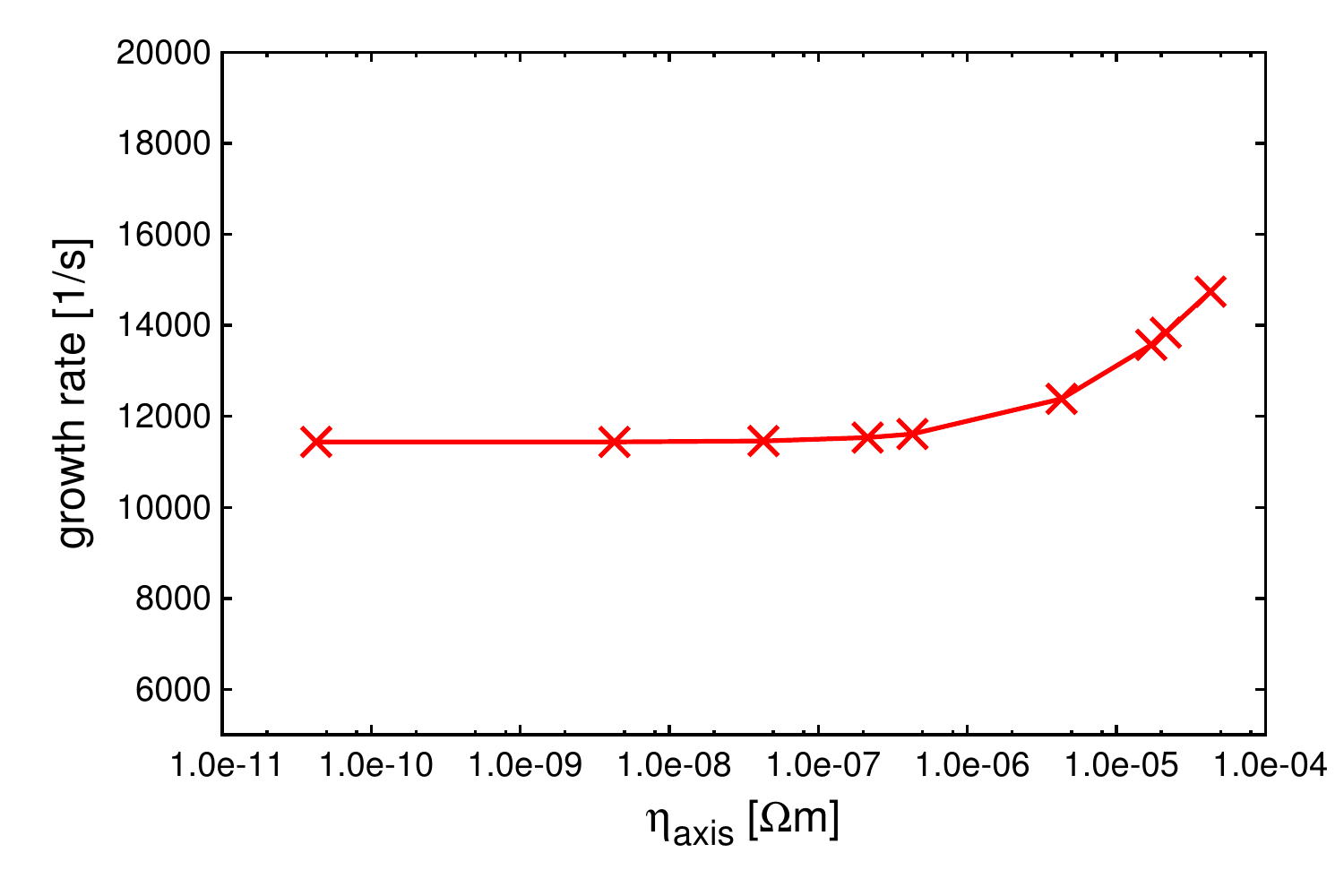}
  \caption{Left: Safety factor profile that has been used for the linear stability analysis. $\Phi_N$ denotes the normalized toroidal magnetic flux. Right: Growth rates of the most unstable mode for different resistivities.}
  \label{fig-linear-equilibrium} 
\end{figure}
 
\begin{figure}
  \centering
  \includegraphics[width=0.4\textwidth]{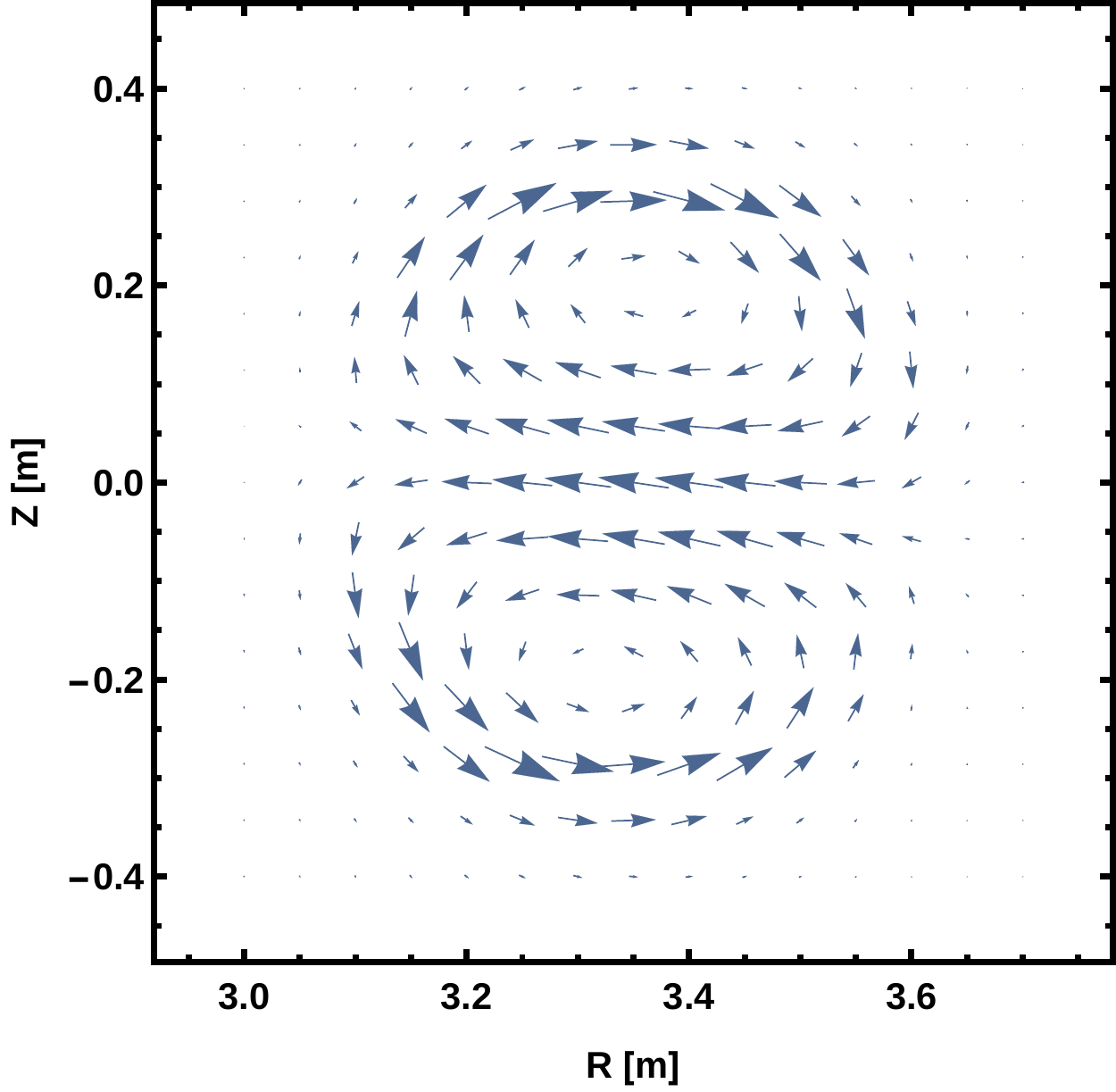}\hspace{4mm}
  \includegraphics[width=0.4\textwidth]{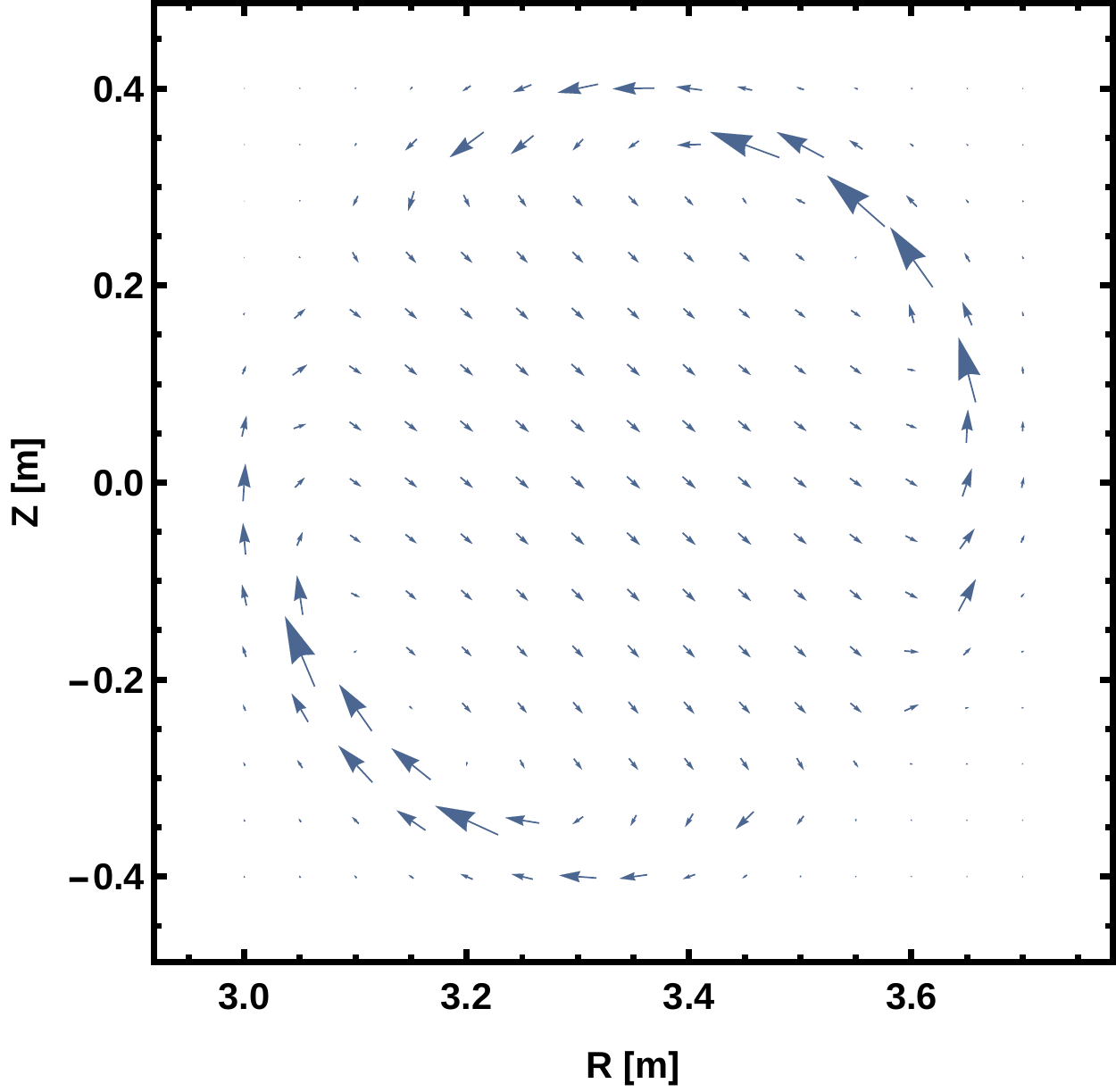}
  \caption{Left: Velocity field in the plasma center for a quasi-interchange instability obtained from the linear stability analysis. Right: For comparison, the velocity field in the plasma center for an internal kink instability is shown.} 
    \label{fig-linear-flow} 
\end{figure}

The calculations have been performed using the linear eigenvalue code {CASTOR3D} \cite{Strumberger2017}. The geometry and parameters are based on a 3D nonlinear simulation (case \lq bb\rq), however, diffusion coefficients have not been included except for the resistivity. As an exact equivalent to the safety factor profile in a 2D equilibrium cannot be calculated from a 3D state, a simple safety factor profile which is flat and close to unity in the plasma center has been chosen (see Figure~\ref{fig-linear-equilibrium}). As expected, it is found that the most unstable mode is ideal as can be seen from Figure~\ref{fig-linear-equilibrium} which shows that its growth rate is approximately independent of the resistivity for a wide range of resistivities. As shown in Figure~\ref{fig-linear-flow}, the mode features the characteristic $(m=1,n=1)$ flow pattern of a quasi-interchange instability which is clearly distinguishable from the flow pattern of an internal kink instability. 

\begin{figure}
  \centering
  \includegraphics[width=0.6\textwidth]{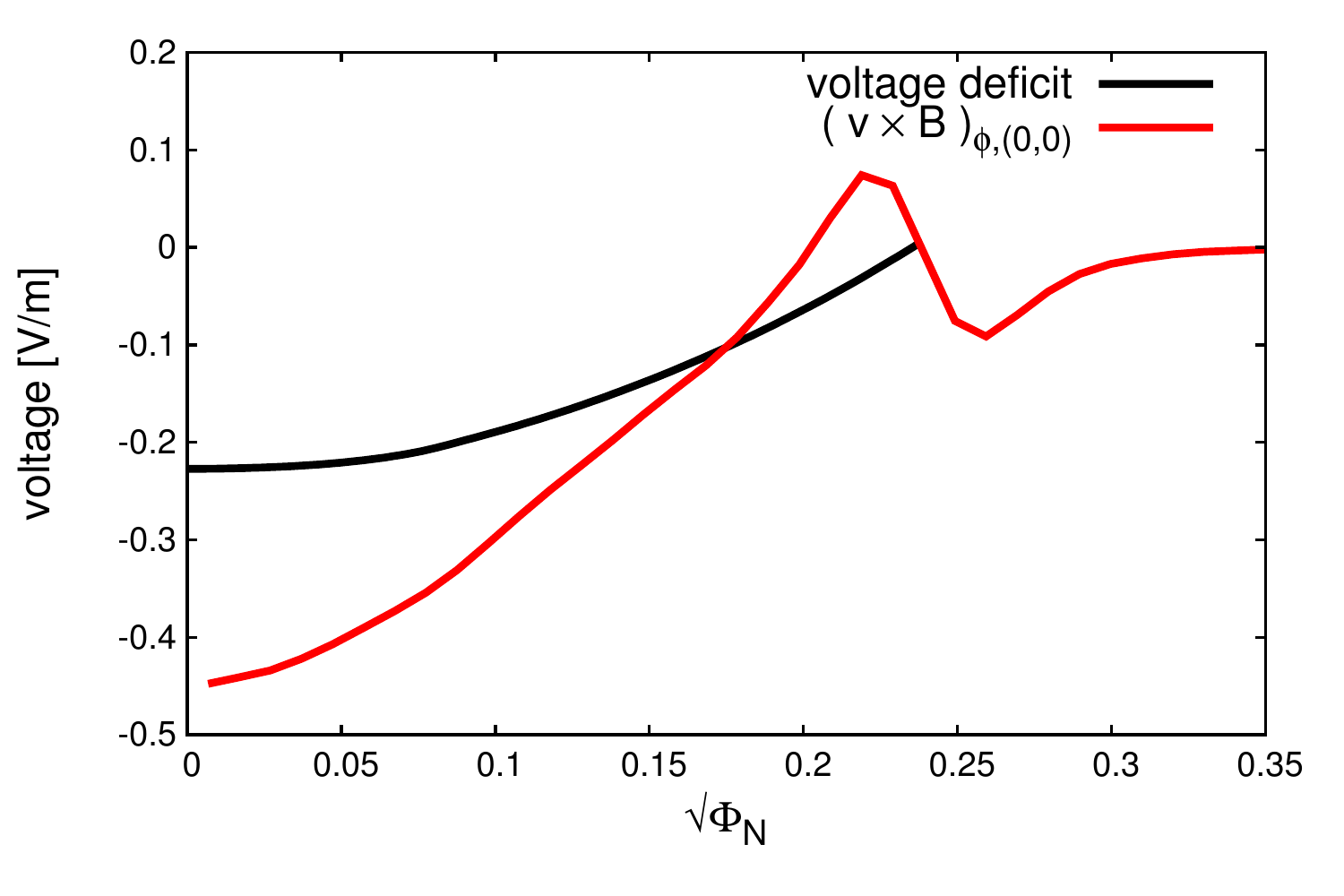}
  \caption{$(m=0,n=0)$ component of the dynamo loop voltage term calculated using the linear eigenfunctions (red) is compared to the voltage that is necessary to raise the safety factor profile in the center to unity in the corresponding 2D nonlinear simulation (black). The linear eigenfunctions have been scaled by a factor such that the maximum of $\sqrt{v^2_R + v^2_Z}$ matches the same quantity in the corresponding 3D nonlinear simulation (case \lq bb\rq, $t= 145600\,\tau_A$).} 
    \label{fig-linear-dynamo} 
\end{figure}

The obtained linear velocity and magnetic field perturbations can be used to calculate a $(m=0,n=0)$ dynamo loop voltage $\left( \mb{v} \times \mb{B} \right)_{\phi}$ in the plasma center as shown in Figure~\ref{fig-linear-dynamo} (red curve). The resulting dynamo loop voltage is compared to the amount of loop voltage that needs to be provided in the 3D nonlinear simulation in order to maintain $q\approx 1$ within the central region of the plasma (black curve). This voltage deficit is calculated from the safety factor profile in the corresponding 2D nonlinear simulation as $2 B_{\phi,2D} \eta_{2D}( 1 - 1/q_{2D}) /(\mu_0 R)$. The comparison shows that the strength of the dynamo loop voltage which has been calculated from the scaled linear eigenfunctions is comparable to the dynamo loop voltage in the 3D nonlinear simulation, even exceeding it by a factor of two in the center. Note, that perfect agreement is not expected as the linear stability calculation is based on a 2D axisymmetric equilibrium whereas the sawtooth-free state in the 3D nonlinear calculation has a 3D helical core. Also, in the 3D nonlinear simulation, the self-regulation mechanism discussed in the previous section can adjust the amount of flux pumping to the voltage deficit. In contrast to this result for the quasi-interchange instability, it is found that a comparable, sufficiently strong dynamo loop voltage cannot be obtained from the linear perturbations of an internal kink instability.

\section{Summary and Outlook}

We have presented a large set of long-term 3D nonlinear MHD simulations in toroidal geometry that have been set up such that a heat source works towards driving the central safety factor to values below unity. The resulting asymptotic states are either characterized by repeating sawtooth-like reconnection cycles or by a stationary flat central safety factor profile with values close to unity. In the sawtooth-free cases, a saturated quasi-interchange instability leads to a stationary helical $(m=1,n=1)$ perturbation of the plasma core, in particular a $(m=1,n=1)$ convection cell. The most important mechanism which prevents $q_0$ from falling significantly below unity in the sawtooth-free states is that this $(m=1,n=1)$ flow combines with the perturbation of the magnetic field to generate an effective negative loop voltage via a dynamo effect as proposed in \cite{Jardin2015}. It is found that the magnetic flux pumping mechanism in the simulations is only able to prevent sawtoothing at sufficiently high $\beta$. Above this threshold, the quasi-interchange instability is sufficiently strong to generate the necessary amount of magnetic flux pumping in order to counterbalance the tendency of the current density profile to centrally peak. It is shown that the dynamo loop voltage mechanism is self-regulating.

A linear stability analysis of an equilibrium with low central magnetic shear and $q_0\approx 1$ confirms that the most unstable mode in this configuration can be characterized as a quasi-interchange instability and that the resulting linear velocity and magnetic field perturbations can be combined to calculate a dynamo loop voltage comparable to the one obtained in the 3D nonlinear simulations.

As a next step, it would need to be examined in more detail if the presented results obtained from the 3D nonlinear MHD simulations can be used to explain the phenomenon of magnetic flux pumping in Hybrid discharges. Corresponding experiments could, e.g., test the existence of a threshold in $\beta$ for flux pumping to occur and that this threshold depends on how low $q_0$ should be according to the applied loop voltage and the amount of on-axis current drive and heating. Furthermore, experimental observations specific to Hybrid discharges in the different tokamaks, like the dependence of flux pumping on the presence of a $(3,2)$ neoclassical tearing mode \cite{Petty2009} or an externally excited $(m=1,n=1)$ perturbation of the plasma core \cite{Taylor2016} in DIII-D, need to be understood in detail. Simulations using an ASDEX Upgrade like geometry are subject of ongoing work. 

\section*{Acknowledgments}
The authors would like to thank D.~Pfefferl\'e, H.~Zohm, V.~Igochine and S.~Hudson for valuable discussions, and J.~Chen for important technical support. Essential software support has been provided by the SCOREC team at RPI. The simulations have been performed on the PPPL cluster. This work was supported by the US DoE Award No. DE-AC02-09CH11466, the Max-Planck/Princeton Center for Plasma Physics, and the SciDAC Center for Extended MHD Modeling.

Notice: This manuscript is based upon work supported by the U.S. Department of Energy, Office of Science, Office of Fusion Energy Sciences, and has been authored by Princeton University under Contract Number DE-AC02-09CH11466 with the U.S. Department of Energy. The publisher, by accepting the article for publication acknowledges, that the United States Government retains a non-exclusive, paid-up, irrevocable, world-wide license to publish or reproduce the published form of this manuscript, or allow others to do so, for United States Government purposes.

\FloatBarrier
\section*{Appendix}
\label{appendix}

In Tables~\ref{tab-basic-param} and \ref{tab-basic-overview}, more details on the simulation set-up are given.
\begin{table}
\renewcommand{\arraystretch}{2}
\begin{center}
  \begin{tabular}{ l l }
  \hline
perpendicular thermal diffusivity & $\chi_{\perp} \approx 5. \cdot 10^{10} \cdot \kappa_0 \cdot T[\si{\kelvin}]^{-1/2} \, \si{\meter^2\per\second}$\\
 & $\hspace{6.6mm} \approx 1.3 \cdot 10^1\,..\,2.3 \cdot 10^4 \si{\meter^2\per\second}$\\
parallel thermal diffusivity & $\chi_{\parallel} \approx 3.45 \cdot 10^7 \, \si{\meter^2\per\second}$\\
energy source & $S = 1.25 \cdot 10^{-17} \cdot \frac{a_S}{d^2_S} \cdot \exp \left( \frac{\left( R - R_{\tn{axis}} \right)^2 + Z^2}{-2.03 \cdot d^2_S}  \right)\, \si{\pascal/\second}$\\
resistivity & $\eta \approx 4. \cdot 10^{-6} \cdot \left( \frac{T}{T_{\tn{axis}}} \right)^{-\frac{3}{2}}\, \si{\ohm \meter}$\\
viscosity & $\nu = 2.3 \cdot 10^{-6} \, \si{\kilogram /(\meter \second)}$\\
toroidal magnetic field on axis & $B_{\tn{axis}} = 1\, \si{\tesla}$\\
target total current for feedback control & $I_{\tn{tot}} = 6.4 \cdot 10^{5}\, \si{\ampere}$\\
target total number of particles for feedback control & $n_{\tn{tot}} = 2.3 \cdot 10^{21}$\\
shape of last closed flux surface & $R [\si{\meter}] = 3.2 + \cos(\theta + 0.2 \sin \theta)$\\
 & $Z [\si{\meter}] = 1.3  \sin \theta$\\
time normalization & $\tau_{A} = 2.90 \cdot 10^{-7} \si{\second}$ \\
 \hline
  \end{tabular}
  \caption{Parameters used for the presented set of simulations. The values for $\kappa_0$, $a_S$ and $d_S$ for the different cases are listed in Table~\ref{tab-basic-overview}.}
  \label{tab-basic-param}
\end{center}
\renewcommand{\arraystretch}{1}
\end{table}

\begin{table} 
\renewcommand{\arraystretch}{1.2}
\begin{center}
  \begin{tabular}{ | c || c | c | c | c | c | c | c | c | c |}
    \hline
    case       & r       & c      & g      & bb     & ee     & cc     & c025   & c05    & c1     \\ \hline
    $\kappa_0$ & 1.75e-6 & 3.5e-6 & 7.e-6  & 5.6e-5 & 1.1e-4 & 1.2e-6 & 3.5e-6 & 3.5e-6 & 3.5e-6 \\ \hline
    $a_S$      & 5.e22   & 1.0e23 & 2.0e23 & 1.6e24 & 3.2e24 & 2.5e22 & 2.2e22 & 3.7e22 & 5.8e22 \\ \hline
    $d_S$      & 0.5     & 0.5    & 0.5    & 0.5    & 0.5    & 0.5    & 0.5    & 0.5    & 0.5    \\ \hline \noalign{\bigskip}
    \hline
    case       & r01     & m3      & m0     & n      & d      & c15    & ff      & wb     & ii      \\ \hline
    $\kappa_0$ & 1.75e-6 & 1.76e-5 & 2.2e-6 & 2.8e-5 & 1.4e-5 & 3.5e-6 & 2.18e-3 & 2.8e-5 & 5.6e-5  \\ \hline
    $a_S$      & 2.3e22  & 8.e22   & 0.0    & 8.e23  & 4.e23  & 8.7e22 & 6.40e25 & 7.4e23 & 1.48e24 \\ \hline
    $d_S$      & 0.5     & 0.5     & --     & 0.5    & 0.5    & 0.5    & 0.5     & 0.5    & 0.5     \\ \hline \noalign{\bigskip}
    \hline
    case       & p     & o     & aa     & qb     & h      & m4n     & z      & mm      & nX3    \\ \hline
    $\kappa_0$ & 5.e-6 & 7.e-6 & 3.5e-6 & 3.5e-6 & 3.5e-6 & 3.52e-5 & 4.4e-6 & 5.5e-6  & 8.4e-5 \\ \hline
    $a_S$      & 8.e22 & 8.e22 & 3.5e22 & 7.1e22 & 8.0e22 & 1.4e23  & 9.5e21 & 1.15e22 & 2.4e24 \\ \hline
    $d_S$      & 0.4   & 0.4   & 0.4    & 0.4    & 0.4    & 0.4     & 0.4    & 0.4     & 0.5    \\ \hline
  \end{tabular}
  \caption{Parameters $\kappa_0$, $a_S$ and $d_S$ as defined in Table~\ref{tab-basic-param} for the different cases. Note that case \lq nX3\rq{ }has a three times larger resistivity.}
  \label{tab-basic-overview}
\end{center}
\renewcommand{\arraystretch}{1.}
\end{table}

The equations that are solved by the M3D-C$^1$ code for the above presented calculations (in SI units) are:
\begin{align}
&\frac{\partial n}{\partial t} + \nabla \cdot \left( n \mb{v} \right) = d_{n} \nabla^2 n + S_{n}&\\
&\frac{\partial \mb{B}}{\partial t} = - \nabla \times \mb{E}&\\
&n m_i \left( \frac{\partial \mb{v}}{\partial t} + \mb{v} \cdot \nabla \mb{v} \right) = - \nabla p + \mb{J} \times \mb{B} + \nu \nabla^2 \mb{v}&\\
&\frac{3}{2} n \frac{\partial T}{\partial t} + \frac{3}{2} n \mb{v} \cdot \nabla T + n T \nabla \cdot \mb{v}& \nonumber \\
&\quad = \nu |\nabla \mb{v}|^2 + \eta \mb{J}^2 + \nabla \cdot \left( \chi_\perp n \nabla T + \chi_\parallel n \frac{\mb{B} \mb{B}}{B^2} \cdot \nabla T \right)  + S\,.&
\end{align}
Here, $n$ is the particle density, $T$ is the sum of the ion and electron temperatures (in $\si{\electronvolt}$), $\mb{v}$ the fluid velocity, $p = nT$ is the total pressure, $S_n$ and $S$ are the particle and energy sources, $\nu$ the dynamic viscosity, $\eta$ the resistivity, $\chi_\perp$ and $\chi_\parallel$ are the perpendicular and parallel heat diffusion coefficients, $d_n$ is an additional anomalous particle diffusion coefficient, and $m_i$ is the ion mass. The magnetic field $\mb{B}$, the electric field $\mb{E}$ and the electric current density $\mb{J}$ are defined as
\begin{align}
&\mb{B} = \nabla \times \mb{A}&\\
&\mb{E} =- \mb{v} \times \mb{B} + \eta \mb{J}&\\
&\mb{J} = \frac{1}{\mu_0} \nabla \times \mb{B}\,,&
\end{align} 
where $\mb{A}$ is the magnetic vector potential.

\bibliographystyle{ieeetr}
\bibliography{database}

\end{document}